 \documentclass[twocolumn,trackchanges]{aastex631}
\shorttitle{Identifying Variable mid and late-T dwarfs}
\shortauthors{Oliveros-Gómez et al.}
\graphicspath{{./}{figures/}}
\usepackage{booktabs}
\usepackage{rotating}
\usepackage{verbatim} 
\usepackage[normalem]{ulem}
\usepackage{longtable}

\begin{document}

\title{Informed systematic method to identify variable mid and late-T dwarfs}

\author[0000-0001-5254-6740]{Natalia Oliveros-Gomez}
\affiliation{Departamento de Astronomía, Universidad de Guanajuato. Callej\'on de Jalisco, S/N, 36023, Guanajuato, GTO, M\'exico}

\author[0000-0003-0192-6887]{Elena Manjavacas}
\affiliation{AURA for the European Space Agency (ESA), ESA Office, Space Telescope Science Institute, 3700 San Martin Drive, Baltimore, MD, 21218 USA}
\affiliation{Department of Physics and Astronomy, Johns Hopkins University, Baltimore, MD 21218, USA}

\author[0000-0003-4912-5229]{Afra Ashraf}
\affiliation{Department of Physics \& Astronomy, Barnard College 
3009 Broadway, New York, NY 10027, USA}
\affiliation{Department of Astrophysics, American Museum of Natural History 
200 Central Park West, New York, NY 10024, USA}

\author[0000-0001-8170-7072]{Daniella C. Bardalez-Gagliuffi}
\affiliation{Department of Physics \& Astronomy, Amherst College, 25 East Drive, Amherst, MA 01003, USA}
\affiliation{Department of Astrophysics, American Museum of Natural History 
200 Central Park West, New York, NY 10024, USA}

\author[0000-0003-0489-1528]{Johanna M. Vos}
\affiliation{Department of Astrophysics, American Museum of Natural History 
200 Central Park West, New York, NY 10024, USA}

\author[0000-0001-6251-0573]{Jacqueline K. Faherty}
\affiliation{Department of Astrophysics, American Museum of Natural History 
200 Central Park West, New York, NY 10024, USA}

\author[0000-0001-7356-6652]{Theodora Karalidi}
\affiliation{Department of Physics, University of Central Florida, 4000 Central Florida Blvd., Orlando, FL 32816, USA}

\author[0000-0001-7356-6652]{Daniel Apai}
\affiliation{Steward Observatory, The University of Arizona, 933 N. Cherry Avenue, Tucson, AZ 85721, USA}
\affiliation{Lunar and Planetary Laboratory, The University of Arizona, 1629 E. University Blvd., Tucson, AZ 85721, USA}



\begin{abstract}

The majority of brown dwarfs show some level of photometric or spectro-photometric variability in different wavelength ranges. This variability allow us to trace the 3D atmospheric structures of variable brown dwarfs and directly-imaged exoplanets with radiative-transfer models and mapping codes. Nevertheless, to date, we do not have an informed method to pre-select the brown dwarfs that might show a higher variability amplitude for a thorough variability study. In this work, we designed and tested near-infrared spectral indices to pre-select the most likely variable mid- and late-T dwarfs, which overlap in effective temperatures with directly-imaged exoplanets. We used  time-resolved near-infrared \textit{Hubble Space Telescope} \textit{Wide Field Camera 3} spectra of a T6.5 dwarf, 2MASS~J22282889--431026, to design our novel spectral indices. We tested these spectral indices on 26 T5.5--T7.5 near-infrared SpeX/IRTF  spectra, and we provided eight new mid- and late-T variable candidates. We estimated the variability fraction of our sample in $\mathrm{38^{+4}_{-30}}$\%, which agrees with the variability fractions provided by \cite{metchev2015weather} for mid-to late-T dwarfs. In addition,  two of the three previously known variables in our SpeX spectra sample are flagged as variable candidates by our indices. Similarly, all seven known non-variable in our sample are flagged as non-variable objects by our indices. These results suggest that our spectral indices might be used to find variable mid- and late-T brown dwarf variables. These indices may be crucial in the future to select cool directly-imaged exoplanets for variability studies.

\end{abstract}



\section{Introduction} \label{sec:intro}

Many brown dwarfs and planetary-mass objects show photometric and spectro-photometric variability (e. g, \citealt{apai2013hst}, \citealt{radigan2014strong}, \citealt{metchev2015weather}, \citealt{zhou2018cloud}, \citealt{Biller2018}, \citealt{vos2019search},  \citealt{manjavacas2019ROSS}, \citealt{Vos2020}, \citealt{Vos2022}). Although other scenarios are possible \citep{tremblin2015fingering, tremblin2016cloudless}, the most likely reason to explain brown dwarf variability is the existence of heterogeneous clouds coverage \citep{apai2013hst} in their atmospheres at various pressure levels. 

Variability studies are a powerful tool for the understanding of the dynamics of substellar atmospheres (e.g. 
\citealt{yang2016extrasolar}, \citealt{apai2017zones}, \citealt{manjavacas2021revealing}), which  overlap in effective temperatures and colors to directly-imaged exoplanets \citep{faherty2016population}. Time-resolved spectrophotometric observations allow us to detect variability at different pressure-levels of the atmospheres of brown dwarfs (e.g. \citealt{yang2016extrasolar}). Using radiative-transfer models, we can provide an interpretation of the heterogeneous cloud layers that are introducing the variability measured, at each pressure level \citep{Saumon_Marley2008}. If we have a long enough monitoring baseline covering at least few rotations of the object, mapping codes like \textit{Stratos} \citep{apai2013hst}, \textit{Aeolus} \citep{karalidi2015aeolus}, \textit{Starry} \citep{2019AJ....157...64L}, or \textit{mapping exoplanets} \citep{cowan2017mapping} can be used to derive 2D or 3D brightness distributions that are consistent with the data and may provide exploratory maps of the atmospheres. Such a thorough analysis can be only performed in variable brown dwarfs. However, to determine if a specific brown dwarf is variable or not,  several hours of photometric or spectrophotometric monitoring need to be spent to obtain a light curve with enough time coverage to determine the variability of the object. So far, the brown dwarfs that have been monitored for variability surveys have been selected blindly, many of them resulting in no detecting any significant variability after expending considerable telescope time. 

In the past years, spectral indices in the near-infrared have been used to identify binary L+T or M+T brown dwarf candidates \citep{burgasser2006method, burgasser2010spex, Bardalez-Gagliuffi2014}. These spectral indices have been found to detect some highly variable brown dwarfs, like 2MASS~J21392676+0220226 \citep{burgasser2006method} or 2MASS~J13243553+6358281 \citep{Looper2007} that did not show any signs of binarity, after high resolution imaging follow-up (\citealt{gagliuffi2015high}, \citealt{burgasser2016orbit}). \cite{manjavacas2019cloud} applied these spectral indices to search for binary candidates in a L, T, and Y-dwarf  spectral library of near-infrared HST/WFC3 data, obtaining eight spectral binary candidates.  \cite{manjavacas2019cloud} found that five out of the eight binary candidates were also flagged as known variables in the literature, suggesting that these spectral indices might be also helpful to search for brown dwarf variables. With this aim, \citet{ashraf2022disentangling} selected which indices to target brown dwarf binaries  would be best at differentiating  L/T transition (L7-T3) variables from the rest of the population. Some of the variability candidates proposed by \citet{ashraf2022disentangling} were recently  confirmed as variables \citep{Vos2022}.

In this paper, we designed a novel set of spectral indices to find variable candidates with mid- and late-T (T5.5-T7.5) spectral types using \textit{Hubble Space Telescope} time-resolved near-infrared spectroscopy of 2MASS J22282889--431026 (2M2228) obtained with the \textit{Wide Field Camera 3} instrument.  2M2228 is a T6.5 brown dwarf \citep{burgasser20032mass},  at a distance of d = 10.64$\pm$0.79~pc \citep{faherty2012brown}. \cite{buenzli2012vertical} estimated a $\mathrm{T_{eff}}$ = 900~K, and a log~g = 4.5-5.0, typical of T-dwarf. 2M2228 is variable with a period of 1.43~hr  and a peak to peak amplitude  of $1.85 \pm 0.07 \%$ in the [4.5] Spitzer channel, $1.85 \pm 0.07 \%$ in the $J$-band,
$2.74 \pm 0.11 \%$ in the $H$-band and $5.3 \pm 0.6 \%$ in $\mathrm{H_2O}$-band \citep{buenzli2012vertical}. \cite{metchev2015weather} measured its variability amplitude in 4.6 \% in the \textit{Spitzer} [3.6] band, and an amplitude of 1.6 \% in the [4.5] channel.  \cite{buenzli2012vertical} measured a phase shift of 89.38$\pm$13.18$^o$ between the $J$- and $H$-band HST/WFC3 light curves, which was interpreted as evidence of large-scale longitudinal-vertical structures. The variability amplitudes and phase-shifts were confirmed by \citet{yang2016extrasolar} in data obtained four years later, which indicates the light curve of 2M2228 is relatively stable in long time scales. 

It is particularly important to design spectral indices for mid and late-T-type objects, since they are generally faint. Thus, variability surveys are more challenging to perform than for L and L/T transition brown dwarfs. The spectral indices presented in this paper might also help to identify variable cool directly-imaged exoplanets in the advent of higher resolution and wavelength coverage \textit{James Webb Space Telescope} near-infrared spectra.

This paper is structured as follows: In Section \ref{sec:observations} we provide the details of the \textit{HST/WFC3} observations for 2M2228. In Section \ref{sec:index-method}, we explain the development of the near-infrared spectral indices for detecting variability in T-type dwarfs. In Section \ref{Sec:Candidate-variables}, we compare our results with previous studies. In Section \ref{sec:robustness} we test the robustness of our spectral indices. In Section \ref{Sec:discussion}, we discuss the variability fraction obtained for our sample, and the physics behind the spectral indices. Finally, in Section \ref{Sec:conclusions}, we summarize our conclusions.

\section{OBSERVATIONS}\label{sec:observations}

The data used for our analysis were obtained on July 7, 2011 with the infrared channel of the \textit{Wide Field Camera 3} (WFC3) on-board the \textit{Hubble Space Telescope} (HST) and its G141 grism, providing spectra covering a wavelength range between 1.10 and 1.70 $\mu$m (GO-12314, GO-70170; PI: D. Apai). These spectra were published in \cite{buenzli2012vertical}. 

The observations were performed using a 256 x 256 pixels subarray of the WFC3 infrared channel, with a field of view of approximately 30×30$''$. A SPARS25 was used 1024 × 1024 pixels readout mode with 11 reads per exposure. Each sub-readout has an exposure time of 22.34 s, after an initial first short read at 0.27 s. Allowing for a maximum exposure number of 27,000 counts which is below the average detector capacity.  The spectra were acquired over six consecutive HST orbits, spanning 8.57~hr, obtaining a total of 54 spectra. In each orbit we obtained 9 spectra with a cadence of 241.99 s. The resolution of the spectra is $\sim$100, and the signal-to-noise ratio was $\sim$124.
\vspace{2mm}

\section{Index Method}\label{sec:index-method}

\subsection{Designing novel spectral indexes}
Spectral indices are parameters that allow us to compare the flux in different spectral ranges within the same spectrum of one object. In this study, we designed  spectral indices using near-infrared HST/WFC3 spectra to find candidate variable mid- and late-T dwarfs using any single-epoch near-infrared spectra, with the aim of facilitating the search for photometric and spectro-photometric variability in mid- and late-T dwarfs.

\begin{figure}
    \centering
    \includegraphics[width=1\linewidth]{ 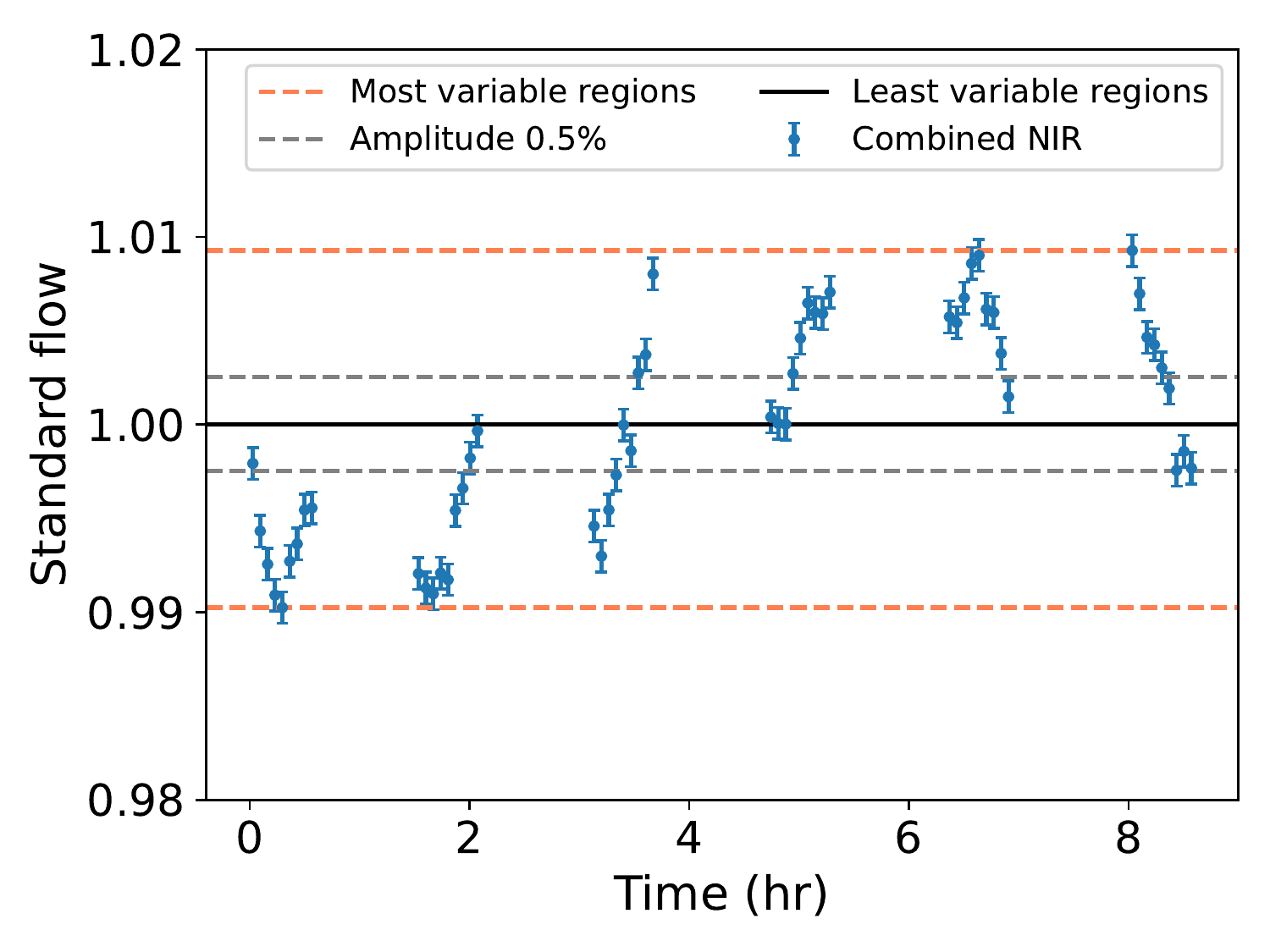}
    \caption{2M2228 near-infrared light curve to identify the regions of variability based on visual inspection. The least variable spectra are those corresponding to the center of the light curve (flux $\sim$ 1,  black line). The most variable spectra are those corresponding to the peaks for the light curve (dashed orange lines).}
    \label{fig:light_curve}
\end{figure}

To design our spectral indices, we used the time-resolved near-infrared HST/WFC3 spectra of 2M2228 published in \citet{buenzli2012vertical}. The near-infrared  light curve of 2M2228 presented in \cite{buenzli2012vertical} showed a variability amplitude between 1.45\% and 5.3\%, depending on the wavelength range of the 2M2228 used to create the light curve (Fig. \ref{fig:light_curve}). In the near-infrared light curve, each data point is a near-infrared time-resolved spectrum itself (see Fig. \ref{fig:light_curve}). To design our spectral indices,  we explored  which wavelength ranges of the HST/WFC3 spectra of 2M2228 vary within the object's rotation due to its heterogeneous cloud-coverage.

\begin{figure}
    \centering
    \includegraphics[width=1\linewidth]{ 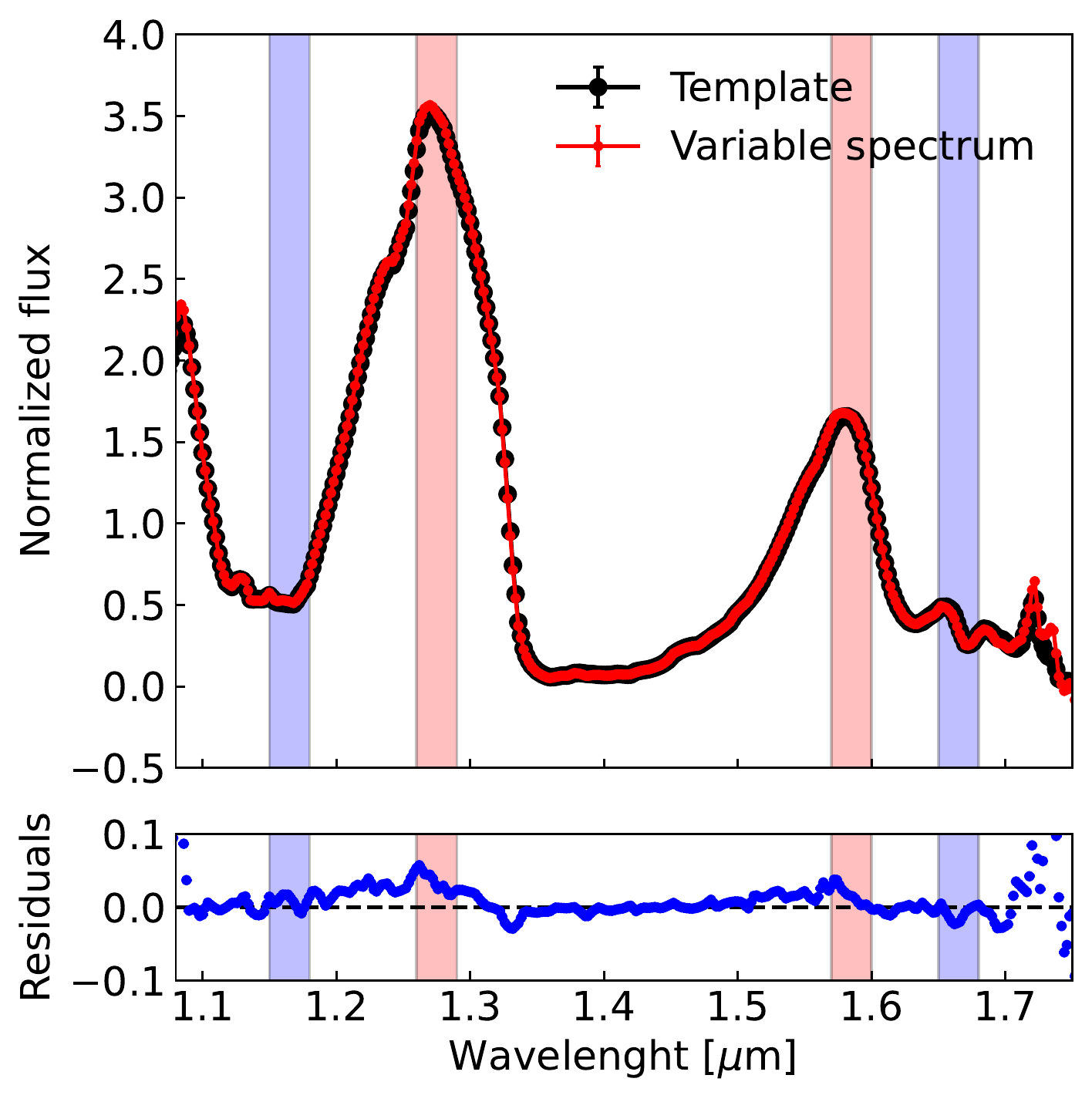}
    \caption{Top panel: template for 2M2228 after median-combining the 40 HST/WFC3 spectra (black), and example of one variable spectrum for 2M2228. Bottom panel: residuals after subtracting the variable spectrum of 2M2228 of the 2M2228 template. We highlight in red the most variable region (the peaks of the the $J$-band and $H$-band), and in blue the least variable regions of the spectrum according to the residuals shown below.}
    \label{fig:residuals}
\end{figure}

\begin{figure*}
\centering
\includegraphics[width=1\linewidth]{ 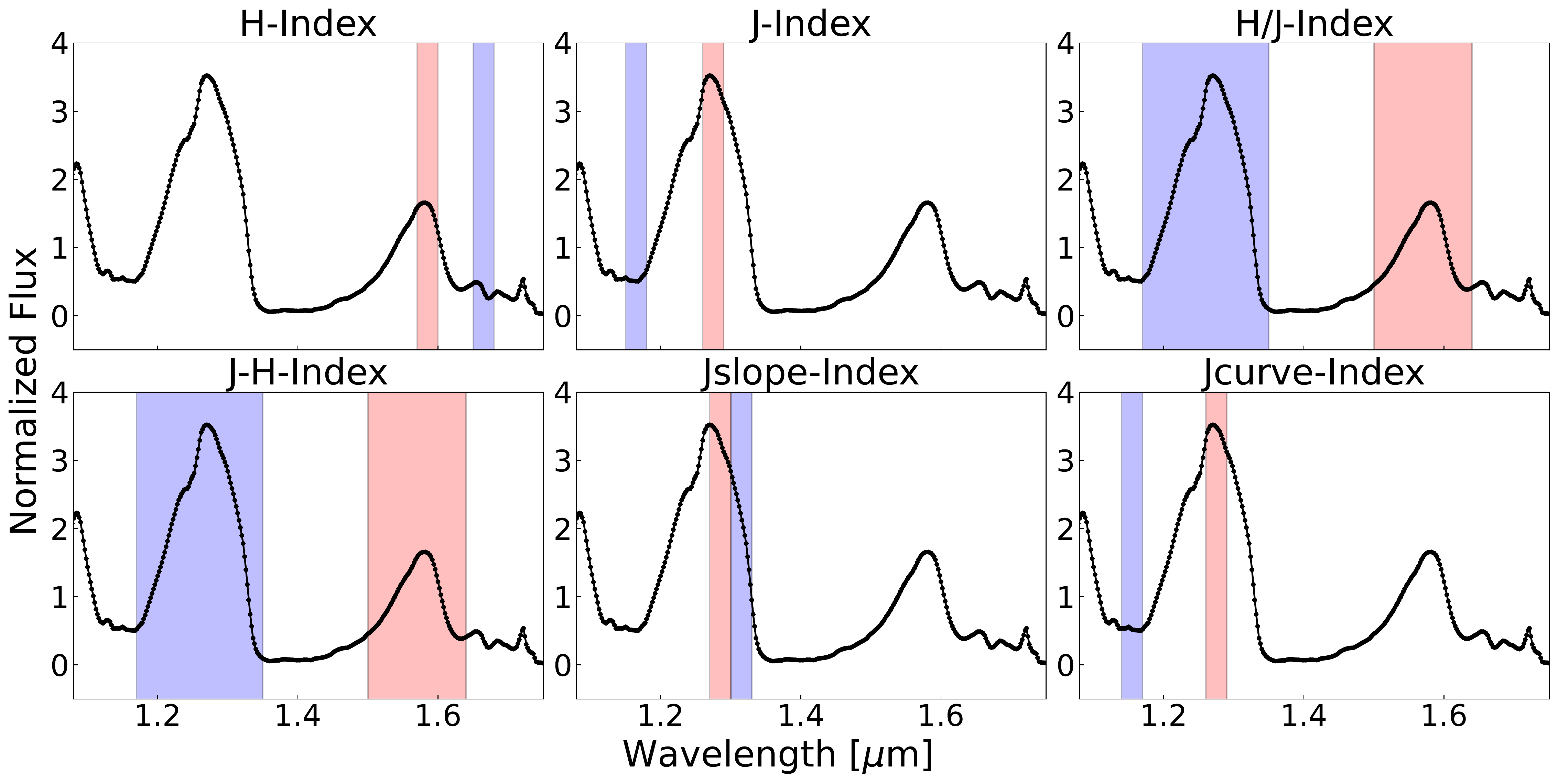}
\caption{Ranges of wavelengths used to calculate spectral indices on a sample brown dwarf spectrum. The red corresponds to the numerator, and the blue corresponds to the denominator in equation \ref{eq:index}.}
\label{fig:ranges_index}
\end{figure*}

We obtained the most and least variable wavelength ranges of the 2M2228 spectrum. For that, we subtracted each of the maximum flux spectra of 2M2228 from a spectral template (Fig. \ref{fig:residuals}), created by median combining all 54 HST/WFC3 2M2228 spectra. Then, we identified the corresponding wavelength ranges in which the difference between the most variable 2M2228 spectrum and the template spectrum is maximum and minimum (in absolute value). The 2M2228 template spectrum would be similar to a single-epoch spectrum of 2M2228 in which several minutes of integration are needed to obtain a high signal-to-noise spectrum. The most variable wavelength ranges are those which have higher absolute difference values, and the least variable wavelength ranges are those that have absolute differences (subtraction) close to zero. We found that the most and least variable regions of each spectrum are similar for all spectra. In Fig. \ref{fig:residuals}, we highlighted in blue the most variable wavelength ranges of the spectra, and in red the least variable wavelength ranges. The wavelength ranges are defined moving average the entire spectrum with 0.03 mm wide windows that are shifted every 0.01 mm. After obtaining the mean values in each window, we selected two regions as the most and least variable. The most variable region show a variability amplitude 1.3 times larger than the overall variability amplitude of the near-infrared light curve. The least variable wavelength range has 0.4 times the variability amplitude of the near-infrared light curve. The least and most variable wavelength ranges are shown in Table \ref{Limits_ranges_index} and Fig. \ref{fig:ranges_index}.

Following this process, we consolidated the spectral ranges to create two spectral indices: one to measure the variability in the $H$-band and the other the $J$-band, which are calculated with equation \ref{eq:index}.

\begin{equation}
    \textup{index} =  \frac{\int_{\lambda_1}^{\lambda_2}F_1(\lambda)d\lambda}{\int_{\lambda_3}^{\lambda_4}F_2(\lambda)d\lambda}
    \label{eq:index}
\end{equation}

In equation \ref{eq:index}, the numerator is one of the most variable region of the spectrum, and the denominator is a adjacent least variable region. Both cover the same wavelength width. In addition, we created two other indices, comparing the $J$- and $H$-bands with each other, one corresponding to the index color, and the other is a ratio between the $H$- and $J$-band, to compare the two most variable bands in our spectrum. Finally, we supplemented our indices with two other indices published in \citet{gagliuffi2014spex} were used $J$-slope and $J$-curve, focusing on the band that has the highest variability, according to the residuals in Fig. \ref{fig:residuals}.  The six indices examined are described in Table \ref{Limits_ranges_index} and shown in Fig. \ref{fig:ranges_index}.

\begin{figure*}
\centering
\includegraphics[width=0.9\linewidth]{ 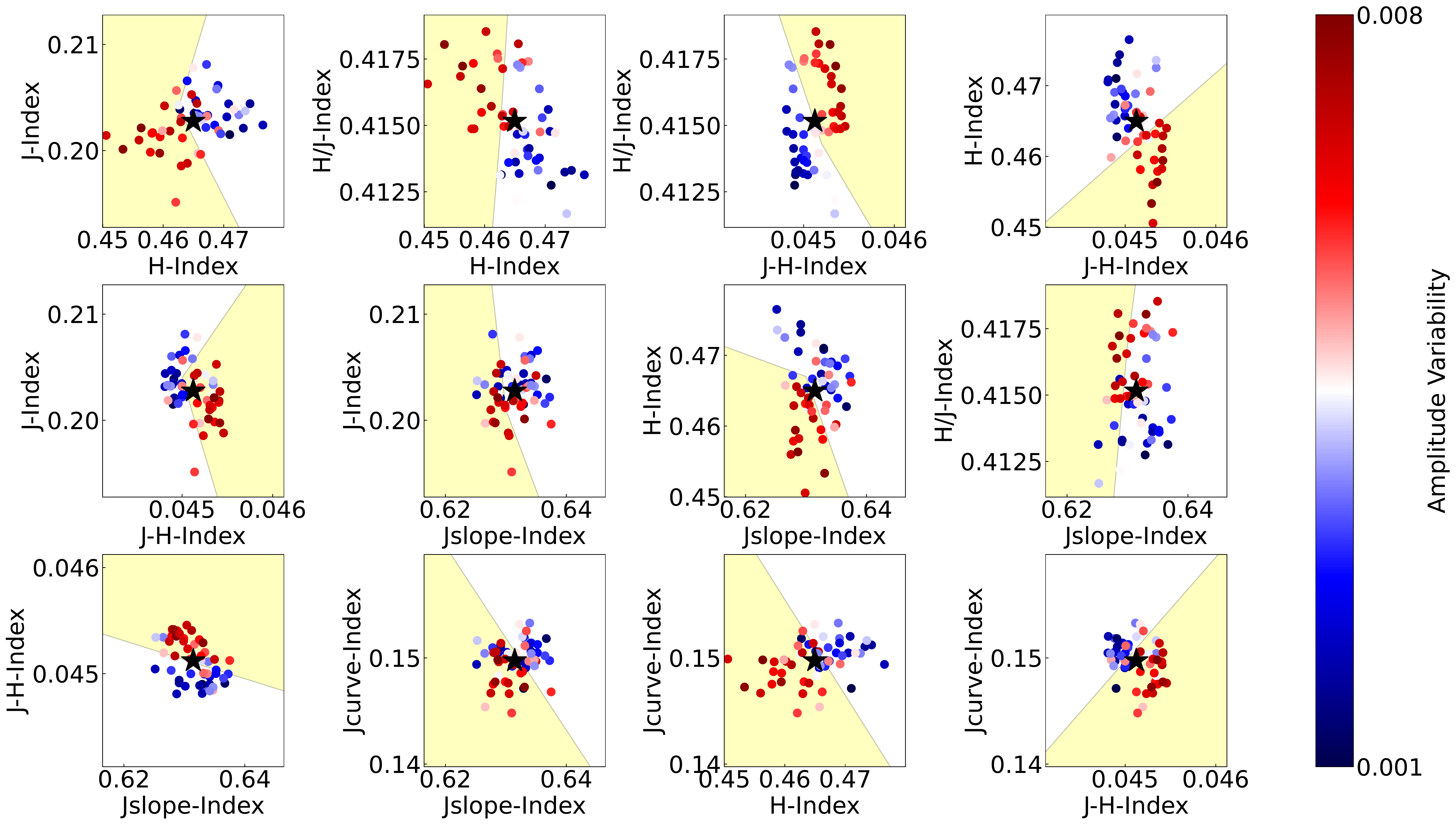}
\caption{Index-index plots comparing the designed spectral indexes to find mid- and late-T brown dwarf variables. The blue points correspond to the spectra that show the least variability amplitude (close to the 1.0 of the light curve in Fig. \ref{fig:light_curve}), and the red correspond to the most variable spectra (close to the 1.007 or 0.992 of the light curve in Fig. \ref{fig:light_curve}). The black
star in the center corresponds to the indexes calculated for the template spectrum with no variability. We show the provisional variable regions (in yellow) using the indices calculated only with observed spectra.}
\label{index-index_comparison}
\end{figure*}

We measured the six spectral indices for the 54 HST/WFC3 NIR spectra of 2M2228. We compared these measurements indices in index-index plots (see Fig. \ref{index-index_comparison}). As observed in Fig. \ref{index-index_comparison}, we see the values of the indices for the most variable and least variable spectra clustering in specific areas of the index-index plots. Among the initial index-index plots, we selected 12 of them that segregated better the most and least variable spectra of 2M2228, and did not resemble any of the other plots. To identify the regions of the index-index plots in which the variable mid- and T-dwarfs would be located, we color coded the points with higher (red) to lower variability amplitudes (blue) spectra, as shown in Fig. \ref{index-index_comparison}.

\subsection{Areas of high variability likelihood in index-index plots}\label{sec:high_like_areas}

The points in Fig. \ref{index-index_comparison} corresponding to the most variable spectra (red points) coincide with the the maximum (or minimum) of the combined near infrared light curve of 2M2228, and the least variable spectra are  those corresponding to the points of the light curve located near a value of 1.0, as shown in Figure \ref{fig:light_curve}. From this analysis, we were able to visually identify the regions where most of the variable spectra are in the index-index plots, as shown in the Fig. \ref{index-index_comparison} (yellow areas). The variability regions presented in Fig. \ref{index-index_comparison}, are limited to the 54 spectra available for 2M2228. Thus, to provide robustly regions of high likelihood of variability in the index-index plots, we generated synthetic spectra, similar to the HST/WFC3 near-infrared spectra, using a Monte Carlo simulation. For that purpose, we redefined each point of the spectrum using a Gaussian random number generator. The mean value of the Gaussian is the original flux in the spectrum, and the standard deviation is the uncertainty of each point of the sample spectrum. 

\begin{figure*}
    \centering
    \includegraphics[width=0.9\linewidth]{ 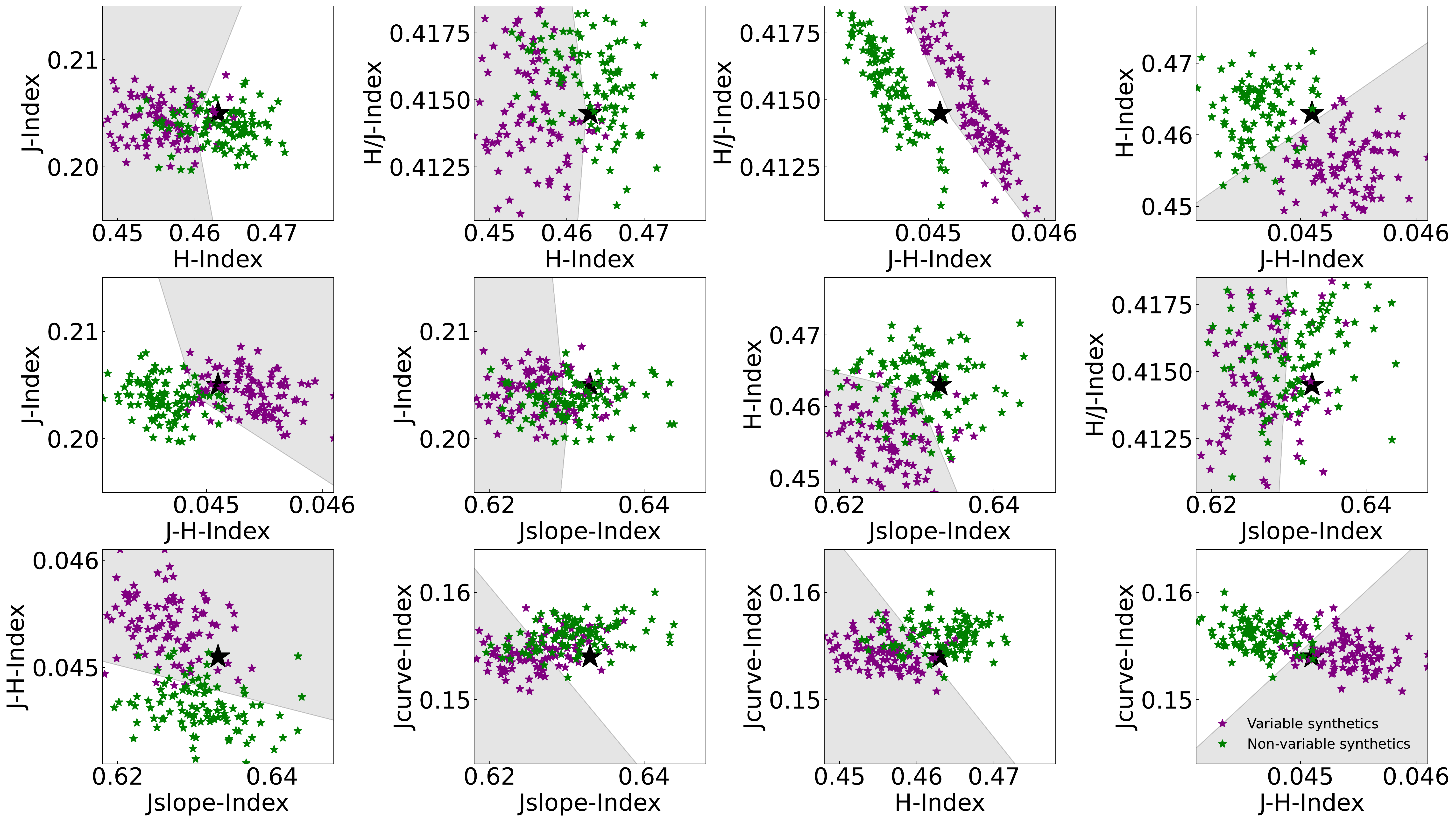}
    \caption{Index-index plot as in Fig. \ref{index-index_comparison} including the value of the indexes calculated for the synthetic spectra created using the most variable spectra of 2M2228 (purple stars), and the least variable 2M2228 spectra (green stars). We show the updated new variable regions (in grey) using the synthetic spectra.}
    \label{fig:regions-synthetic}
\end{figure*}

We generated two different sets of synthetic spectra: the first set similar to the most variable spectrum in the 2M2228 light curve (in the light curve in Fig. \ref{fig:light_curve} at 0.3~hr, normalized flux $\approx$ 0.99024), and second set from the least variable spectrum in the object's light curve. The least variable spectrum of 2M2228 is the one with relative flux value closest to 1.0 in the target's light curve (in Fig. \ref{fig:light_curve} at 3.4~hr, normalized flux = 0.99997). 
We generated 100 new synthetic spectra using the most variable spectrum of 2M2228, and 100 new synthetic spectra using the least variable spectrum of 2M2228. With these 200 new synthetic spectra, the six indices are calculated for each spectrum and plotted alongside the initial 54 spectra of 2M2228 in the index-index comparison plots (Fig. \ref{fig:regions-synthetic}). Using the newly generated synthetic spectra, we  refine the initial variable areas in the index-index plots. The new regions of variability are shown in Fig. \ref{fig:regions-synthetic} and the values for the polygons' limits are described in Table \ref{poligone-areas}.

\subsection{Candidate selection}

\begin{figure}
    \centering
    \includegraphics[width=1\linewidth]{ 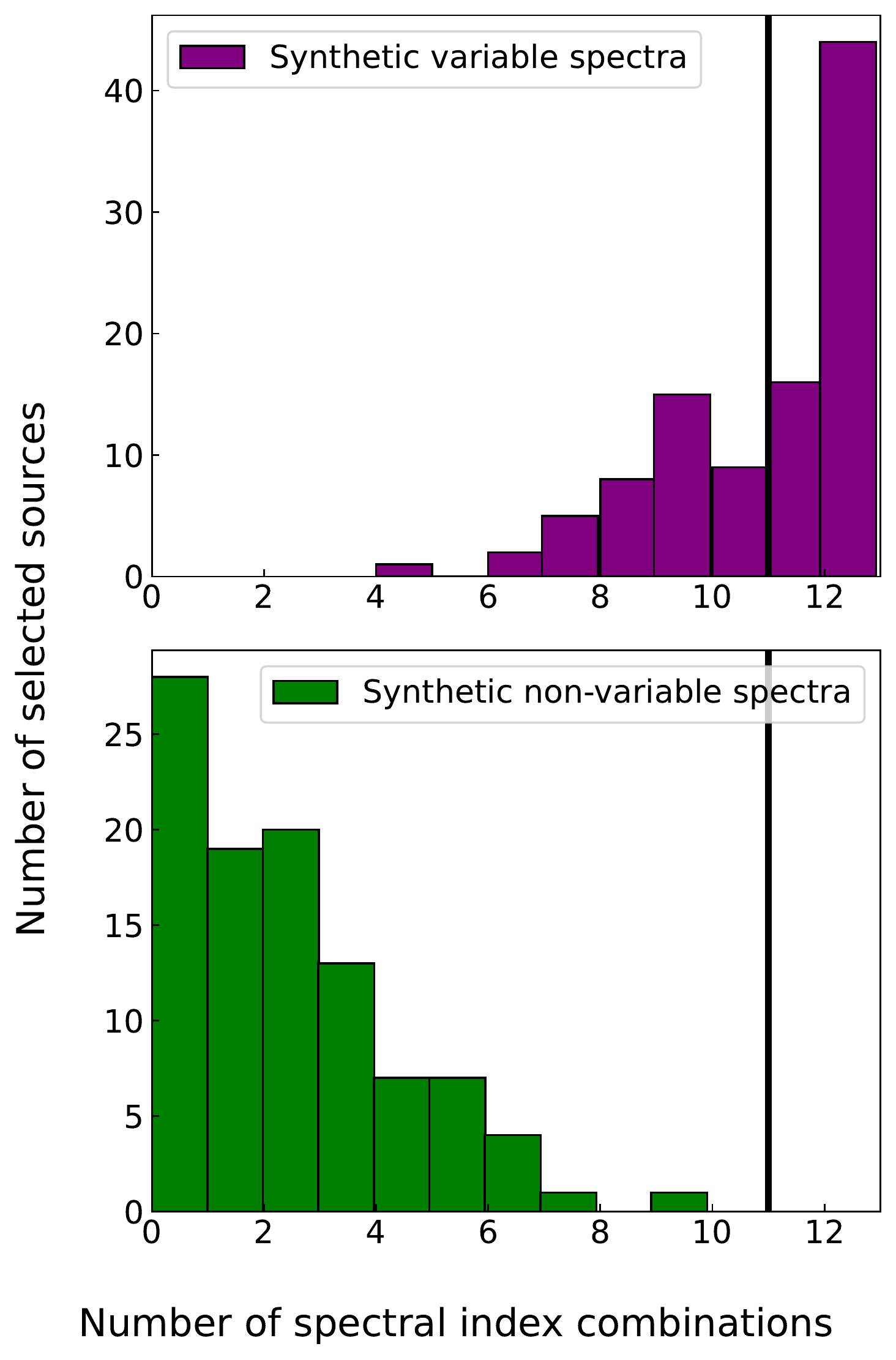}
    \caption{Upper plot: Histogram showing how many synthetic variable spectra fall in variable areas of index-index plots. The histogram shows that most of the variable synthetic spectra fall at least in 11 variable areas in the index-index plots. Bottom plot: Histogram showing how many synthetic non-variable spectra fall in variable areas. The histogram shows that most of the non-variable synthetic spectra fall in less than 10 variable areas. Thus, we mark the threshold on 11 plots to catalog an object as a variable candidate. }
    \label{fig:histogram-synthetic}
\end{figure}

Using the areas defined in Section \ref{sec:high_like_areas}, it is possible to denote variability candidacy based on the number of times a spectrum is found within the regions of variability. Figure \ref{fig:histogram-synthetic} shows two histograms with the frequency in which spectra are identified as variable or non-variable in a given number of index plots. We found that 70\% of the synthetic spectra generated using the most variable 2M2228 spectrum are within the variable regions in at least 11 of the 12 index-index plots (upper panel). For the case of  the synthetic spectra generated using the least variable 2M2228 spectrum, we found that 100\% of the spectra fall in the variable region in less than 9 of the index-index plots (lower panel). For this reason, we identify as \textit{variable candidates} those found at least in 11 regions in our index-index comparison plots.

\begin{table*}
\centering
\begin{tabular}{ccccc}
\hline
\hline
Spectral & Numerator Range\footnote{Color index case, numerator is the first value.} & Denominator Range\footnote{Color index case, denominator is the second value.} & Feature  &   Reference        \\                    index &     ($\mu$m) & ($\mu$m) & &
\\
\hline
H/J              & 1,51-1,62        & 1,205-1,315          & Comparison of  H-band and J-band    &     (1)                    \\
J-H              & 1,205-1,315      & 1,51-1,62            & J-H Color index        &        (1)                                 \\
H                & 1,59-1,62        & 1,64-1,67            & Difference variability in H-band  & (1)   \\
J                & 1,22-1,25        & 1,15-1,18            & Difference variability in J-band & (1)   \\
$J$-slope           & 1.27-1.30        & 1.30-1.33            & 1.28 $\mu$m flux peak shape   &      (2)                                \\
$J$-curve           & 1.26-1.29        & 1.14-1.17            & Curvature across  J-band  & (2)   \\   
\midrule[0.6mm] 
\end{tabular}
\caption{Boundary ranges used for the creation of the spectral indices. References: (1) This paper, (2) \citet{gagliuffi2014spex} \label{Limits_ranges_index}}
\end{table*}

\begin{table*}
\centering
\begin{tabular}{c c c}
\hline
\hline

x-axis  & y-axis  & Coordinates                                                                   \\ \hline
H       & J       & (0.445,0.215), (0.46,0.215), (0.463,0.2045), (0.475,0.19), (0.445,0.19)  \\ 
H       & H/J     & (0.445,0.4185), (0.457,0.4185), (0.464,0.4146), (0.46,0.410), (0.445,0.410) \\ 
J-H     & H/J     &    (0.04375,0.422), (0.0465,0.422), (0.0465,0.410), (0.0459,0.410),(0.045,0.4146)             \\ 
J-H     & H       & (0.0465,0.485), (0.0465,0.445), (0.044,0.445)  \\ 
J-H     & J       & (0.0447,0.215), (0.04615,0.215), (0.04615,0.19), (0.045,0.2045) \\ 
J-slope & J       & (0.612,0.215), (0.6275,0.215), (0.632,0.2042), (0.6275,0.1915), (0.612,0.1915) \\ 
J-slope & H       & (0.612,0.47), (0.632,0.463), (0.646,0.448), (0.612,0.448)          \\ 
J-slope & H/J     & (0.612,0.4186), (0.6275,0.4186), (0.632,0.4146), (0.6275,0.4105), (0.612,0.4105) \\ 
J-slope & J-H     & (0.612,0.046), (0.65,0.046), (0.65,0.0447), (0.612,0.04515)   \\ 
J-slope & J-curve & (0.613,0.165), (0.653,0.142), (0.615,0.142) \\ 
H       & J-curve & (0.447,0.164), (0.481,0.142), (0.447,0.142) \\ 
J-H     & J-curve & (0.0462,0.164), (0.0462,0.142), (0.0438,0.142)  \\ \midrule[0.6mm]
\end{tabular}
\caption{Boundaries for the  variable areas within the index-index plots in Fig. \ref{index-index_comparison}.}
\label{poligone-areas}
\end{table*}

\section{IDENTIFICATION OF CANDIDATE VARIABLES}\label{Sec:Candidate-variables}

Once robust areas were defined within the index-index plots, we are able to use our spectral indices to identify variable mid- to late-T brown dwarf candidates. 

We used all available near-infrared (0.7--2.5~$\mu$m) spectra from the  SpeX/IRTF spectral library\footnote{\url{ http://pono.ucsd.edu/~adam/browndwarfs/spexprism/html/tdwarf.html}} (R$\sim$120) with spectral types between T5.5 and T7.5. We end up with 26 brown dwarfs with a single epoch spectrum. We  test our spectral indices with brown dwarfs with spectral types between T5.5 and T7.5 since they are within one spectral type of 2M2228 (T6.5). In this sample, we included the SpeX spectrum of 2M2228 itself to confirm the validity with our method with spectra taken with other instruments different from WFC3.

In Table \ref{table:sample_testing}, we show the details of the T5.5-T7.5 sample that we used to test our method. We calculated the indices for the 26 brown dwarfs, and created the index-index plots, concluding that 10 of the 26 T5.5-T7.5 dwarfs were selected by the indices as variable candidates. This represents a $\sim$38\% variability occurrence for the T5.5 - T7.5 spectral sample considered in this study.

\begin{figure}
    \centering
    \includegraphics[width=1\linewidth]{ 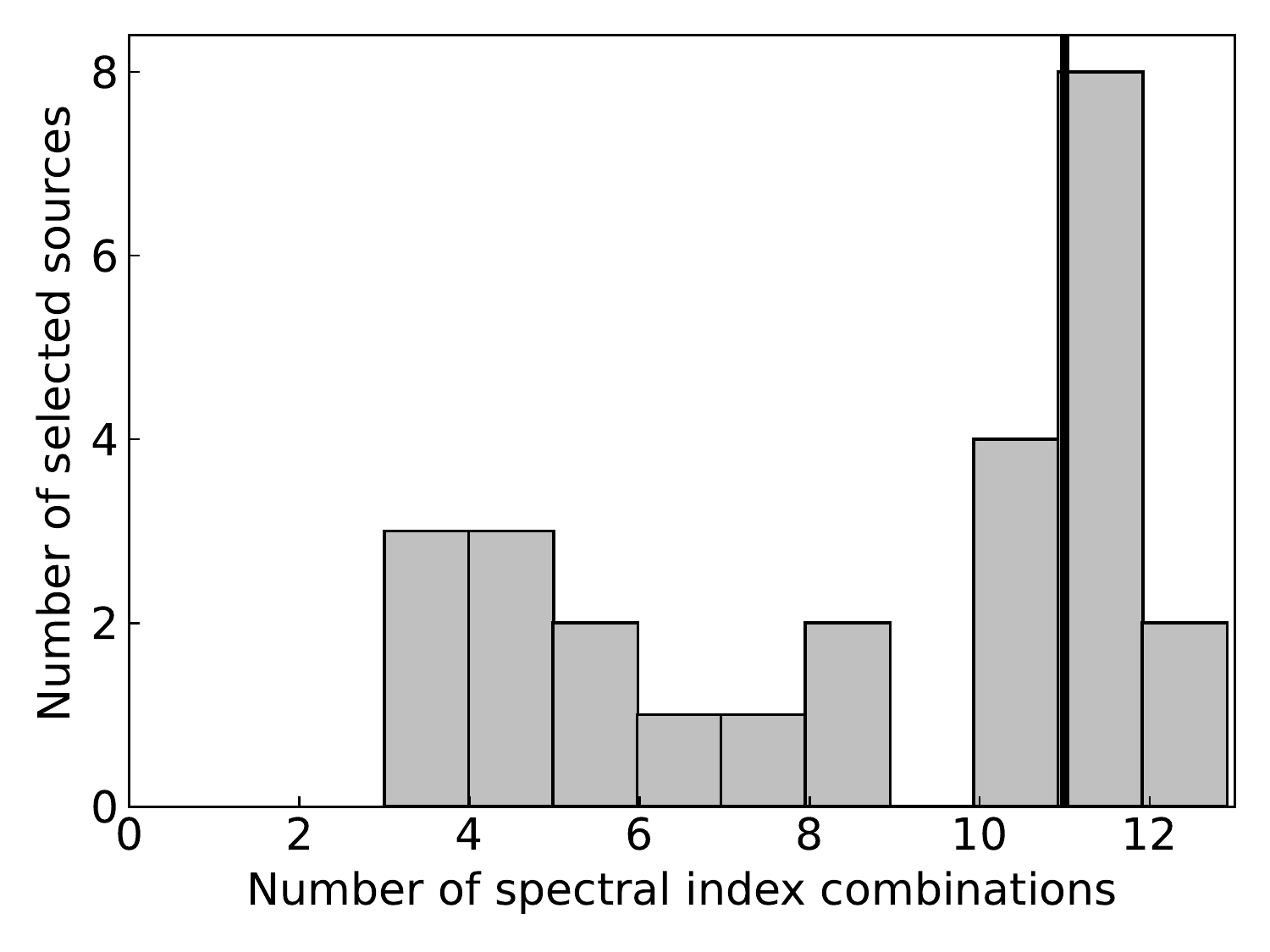}
    \caption{Histogram of the number of times each of the mid- to late-T dwarfs  in  Table \ref{table:sample_testing} fall in the variability regions in  each of the 12 index-index plots. }
    \label{fig:AllBrownDwarf}
\end{figure}

\begin{table*}
\centering
\caption{T dwarfs with spectral type between T5.5-7.5 available at the SpeX Spectral Library. Variability references: (1) \cite{koen2004search}, (2) \cite{buenzli2014brown}, (3) \cite{radigan2014strong}, (4) \citet{allers2020measurement}, (5) \cite{dupuy2012hawaii},
(6) \cite{metchev2015weather}, (7) \cite{clarke2008search}, (8) \cite{vos2018variability}, (9) \cite{buenzli2012vertical}}
\label{table:sample_testing}
\begin{turn}{90}
\begin{tabular}{cccccccccc}
\hline
\hline
Name                          & RA          & DEC          & SpeX NIR & $J$-mag     & Our variable & N° & Variability & Reference \\
& & & SpT & &  candidate & regions &   literature &  variability studies \\
\hline

2MASSI   J0243137--245329      & 02 43 13.71 & -24 53 29.83 & T6.0      & $15,381 \pm 0,05$   & No                   & 04/12           & Non-variable  &    (1)(2)(3)       \\
2MASSI   J0727182+171001      & 07 27 18.24 & 17 10 01.20  & T7.0      & $15.600 \pm 0.061$  & Yes                  & 11/12             & 
 &                \\
2MASSI J0937347+293142        & 09 37 34.87 & 29 31 40.99  & T6.0      & $14.648 \pm 0.036$  & No                   & 05/12            &    &               \\
2MASSI J1047538+212423
& 10 47 53.85 & 21 24 23.46  & T6.5    & $15.819 \pm 0.059$  & No                   & 10/12            &   Variable &   (4)            \\
2MASSI J1217110--031113        & 12 17 11.10 & -03 11 13.17 & T7.5    & $15.860 \pm 0.061$  & Yes                  & 11/12             &   &                \\
2MASSI J1534498--295227        & 15 34 49.84 & -29 52 27.41 & T5.5    & $ 14.900 \pm 0.054$ & No                   & 03/12             &   &                \\
2MASSI J1553022+153236        & 15 53 02.28 & 15 32 36.91  & T7.0      & $15.825 \pm 0.071$  & Yes                  & 11/12             & Binary  &  (5)       \\
2MASSI J2356547--155310        & 23 56 54.77 & -15 53 11.13 & T5.5    & $15.824 \pm 0.057$  & No                   & 03/12             &   &                \\
2MASS J003451+052305       & 00 34 51.57 & 05 23 05.09  & T6.5    & $15.140 \pm 0.004$  & Yes                  & 11/12             &   &                \\
2MASS J051609--044549       & 05 16 09.45 & -04 45 49.93 & T5.5    & $15.984 \pm 0.079$  & No                   & 04/12           & Non-variable   &   (6)            \\
2MASS J111451--261823       & 11 14 51.33 & -26 18 23.56 & T7.5    & $15.858 \pm 0.083$  & No                   & 10/12             &  &                 \\
2MASS J122554--273946       & 12 25 54.32 & -27 39 46.67 & T6.0      & $15.260 \pm 0.047$  & No                   & 08/12             &  &                 \\
2MASS J123147+084733 & 12 31 47.53 & 08 47 33.11  & T5.5    & $15.570 \pm 0.072$  & No                   & 04/12           &   &                \\
2MASS J123739+652614       & 12 37 39.19 & 65 26 14.80  & T6.5    & $16.053 \pm 0.087$  & No                   & 08/12             &   &                \\
2MASS J154627--332511       & 15 46 27.18 & -33 25 11.18 & T5.5    & $15.631 \pm 0.051$  & No                   & 03/12             & Non-variable  & (3)(7)      \\
2MASS J161504+134007       & 16 15 04.13 & 13 40 07.92  & T6.0      & $16.350 \pm 0.092$  & Yes                  & 12/12             &   &                \\
2MASS J182835--484904       & 18 28 35.72 & -48 49 04.63 & T5.5    & $15.175 \pm 0.056$  & No                   & 06/12             & Non-variable  &   (3)(7)              \\
Gliese 570D                   & 14 57 14.96 & -21 21 47.75 & T7.5    & $15.324 \pm 0.048$  & Yes                  & 11/12             &   &                \\
HD 3651B                      & 00 39 18.91 & 21 15 16.80  & T7.5    & $16.31 \pm 0.03$    & No                   & 10/12             &   &                \\
SDSS J175805+463311      & 17 58 05.45 & 46 33 09.91  & T6.5    & $16.152 \pm 0.088$  & Yes                  & 11/12             &   &                \\
SDSSp J111010+011613     & 11 10 10.01 & 01 16 13.08  & T5.5    & $16.161 \pm 0.008$  & No                   & 07/12            & Non-variable   &      (8)         \\
SDSSp J134646--003150     & 13 46 46.34 & -00 31 50.13 & T6.5    & $16.000 \pm 0.102$  & Yes                  & 11/12             &   &                \\
SDSSp J162414+002915     & 16 24 14.36 & 00 29 15.81  & T6 &      $15.494 \pm 0.054$  & No                   & 05/12            & Non-variable  &     (1)(3)           \\
2MASS J005019--332240       & 00 50 19.94 & -33 22 40.27 & T7.0      & $15.928 \pm 0.070$  & Yes                  & 11/12             & Variable &    (3) (6)             \\
ULAS J141623+134836~B      & 14 16 23.94 & 13 48 36.32  & T7.5    & $17.259 \pm 0.017$  & No                   & 10/12             & Non-variable & (6)\\
2MASS J222828--431026       & 22 28 28.89 & -43 10 26.27 & T6.5    & $15.662 \pm 0.073$  & Yes                  & 12/12             & Variable & (6)(9)                \\
\midrule[0.6mm]
\end{tabular}
\end{turn}
\end{table*}

From the SpeX sample, ten objects have been monitored for variability in previous works (\citealt{radigan2014strong}, \citealt{metchev2015weather}). The only known variables, apart from 2M2228 are J0050--3322 and J1047+2124. There are seven others of the T5.5-T7.5 sample that have been confirmed as non-variable or with amplitudes below noise level in previous works \citep{clarke2008search, radigan2014strong, buenzli2014brown, metchev2015weather}.

\begin{figure}
    \centering
    \includegraphics[width=0.8\linewidth]{ 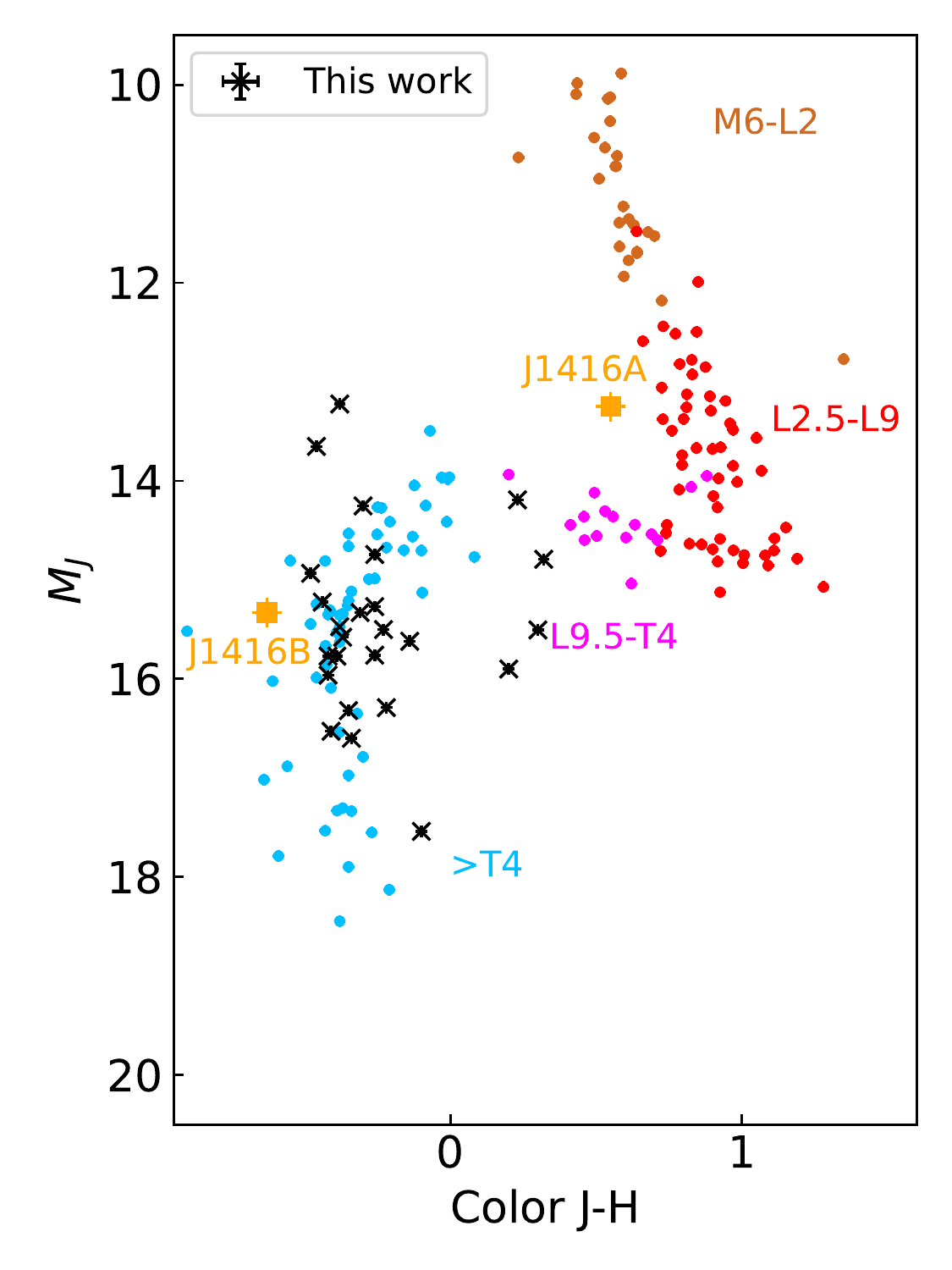}
    \caption{Color–magnitude diagrams in the MKO system showing all ultracool dwarfs from Table \ref{table:sample_testing} in \citet{dupuy2012hawaii} with errors $<0.5$ mag.  Symbol colors indicate spectral types: M6–L2 (brown), L2.5–L9 (red), L9.5–T4 (purple), $>$T4.5 (blue), and the brown dwarfs with IRTF/SpeX spectra used in this work (black). J1416B is one of the objects included in our sample of weak variables that have a  blue color. }
    \label{fig:color-mag}
\end{figure}

Since our spectral indices are designed in the $J$- and $H$-band, there are expected to be applicable to brown dwarfs with similar colors according to the main sequence of mid- and late-T dwarfs. In Fig. \ref{fig:color-mag}, we show the 26 mid- to late-T dwarfs  used to test our indices.

In the following section, we list the objects  selected as variable and non-variable candidates by our indices. The full list of variable candidates selected by our indices can be found in Table~\ref{table:sample_testing}.

\subsection{Variables candidates}\label{sec:variables}

The following objects were selected by our indices as variable candidates.

\subsubsection{2MASS J22282889--431026}\label{2MASS J22282889-431026}

2M2228 was classified as a T6.5 spectral type \citep{burgasser20032mass}. \cite{buenzli2012vertical} measured a variability amplitude of 1.85$\pm$0.07~\% in the $J$-band, and a 2.74$\pm$0.11\% in the $H$-band using near-infrared HST/WFC3 spectra. The $J$-, $H$-band and $\mathrm{H_{2}O}$ band at 1.4~$\mu$m light curves show phase shifts. \cite{buenzli2012vertical} found that the phase lag increases with decreasing pressure level, or higher altitude, probing the heterogeneity in the atmosphere of 2M2228 in both horizontal and vertical directions. 

To test our spectral indices, we used a single-epoch SpeX/IRTF spectrum of 2M2228 from \cite{burgasser20042mass}. 2M2228 was found in the variable areas in 12 out of 12 our index-index plots, validating the effectiveness of our indices. 

\subsubsection{2MASS J00501994--3322402}\label{2MASS J00501994 - 3322402}

J0050--3322 is classified as a T7.0 brown dwarf \citep{kirkpatrick2011first}. Using a forward modelling analysis, \citet{zhang2021uniform} measured an effective temperature of $\mathrm{T_{eff}} = 947^{+23}_{-24}K$, surface gravity log~g =  $4.06^{+0.26}_{-0.26}$, and metallicity of, Z~=~-0.11$^{+0.15}_{-0.15}$. J0050-3322 was monitored by \cite{metchev2015weather} with  \textit{Spitzer}, finding  no significant variability  in the  [3.6] channel, and a variability amplitude of $1.07\pm0.11$\% in the [4.5]  channel.  J0050--3322 have a regular light curve, with a period  of $P = 1.55 \pm 0.02$~hr, being one of the fastest brown dwarf rotators. J0050--3322 was found in the variable areas in 11 out of 12 our index-index plots.

\subsection{Non-variables candidates}\label{sec:non-variables}

The following objects were selected as non-variable candidates by our spectral indices.

\subsubsection{2MASSI J1047538+212423}\label{2MASSI J1047538+212423}

J1047+2124 is a T6.5 dwarf \citep{burgasser2000discovery}. \citet{filippazzo2015fundamental} provided physical parameters for this object: $\mathrm{T_{eff}} = 880 \pm 76$ K, $log(g) = 4.96 \pm 0.49$ dex, age around $0.5-10$ Gyr, M= $41.61 \pm 26.03~\mathrm{M_{jup}}$ and R = $4.96 \pm 0.49~\mathrm{R_{jup}}$. \citet{allers2020measurement} used the Infrared Array Camera on the Spitzer Space Telescope used the Infrared Array Camera on the Spitzer Space Telescope, on its [4.5] channel, for two days, during 7~hr and 14~hr each. \citet{allers2020measurement} found sinusoidal variability of  $0.5\%$, and adopt a period of $1.741\pm0.007$ hr.

In addition, \citet{allers2020measurement} measured the wind speed for this object, in radio observation taken with Karl G. Jansky Very Large Array (VLA), which generates an (unknown) atmospheric inhomogeneity that dominates the photometric variability, with similar period, $1.758 \pm 0.0012$ hr. J1047+2124 was found in the variable areas in 10 out of 12 our index-index plots, indicating that it is  non-variable candidate. Since J1047+2124 is a false negative, in section \ref{sec:robustness}, we analyze in detail this particular object.

\subsubsection{ULAS J141623.94+134836.3 B}\label{ULAS J141623.94 + 134836.3}

J1416+1348 is a resolved L+T subdwarf system, separated by  9''  \citep{burgasser2010spex}. The primary dwarf is ULAS J141623.94+134836.3 (J1416+1348A) with a L7 spectral type. On the other hand, the secondary object J1416+1348B, is a T7.5 dwarf. \citet{burgasser2000discovery} noted the unusually blue $H-W2$ and $W1-W2$ colors of this object and connected these properties to its metal-poor atmosphere. The very blue $Y-K$ color of this companion suggests high gravity and low metallicity, with $logg = 5.21^{+0.26}_{-0.25}$ dex and metallicity $Z=-0.39^{+0.14}_{-0.11}$. \citet{zhang2021uniform} derived an effective temperature of $\mathrm{T_{eff}} = 695 \pm 33$ K. 

Variability studies have been performed on both objects. J1416+1348B was initially studied in \citet{khandrika2013search} and marginal evidence of $J$-band variability was found using the Gemini infrared camera on the 3~m Shane telescope at Lick Observatory.  \citet{radigan2014strong} and \citet{metchev2015weather} found no significant variability. \citet{metchev2015weather}  provided an upper levels of variability of A[3.6] $<$0.91~\% and  A[4.5] $<$0.59~\%. J1416+1348B was found in the variable areas in 10 out of 12 our index-index plots, indicating that it is a non-variable candidate.

\subsubsection{2MASS J05160945--0445499}\label{2MASS J05160945 - 0445499}
2MJ0516--0445 is a T5.5 dwarf, and it is at a parallactic distance of $12 \pm 2$ pc. \citet{faherty2012brown} measured its effective temperature $\mathrm{T_{eff}} = 1211 \pm 137$ K, $log(g) =5.02 \pm 0.48$ dex, and reported an age between $0.5-10$ Gyr.   \citet{faherty2012brown} estimated a mass of M= $47.18 \pm 27.53  ~\mathrm{M_{jup}}$ and R = $ 0.97 \pm 0.17~\mathrm{R_{jup}}$. \citet{metchev2015weather} measured its variability with the \textit{Spitzer} telescope, finding no significant variability, but providing  upper limits on its variability of  A[3.6] $<$ 0.83~\% and  A[4.5] $<$ 0.81~\%. J0516--04454 was found in the variable areas in 4 out of 12 our index-index plots, indicating that it is a non-variable candidate.

\subsubsection{SDSSp J162414.37+002915.6}\label{SDSSp J162414.37+002915.6}

J1624+0029 is a T6.0 dwarf \citep{burgasser2002spectra}.  \citet{smart2013nparsec} provided the physical parameters for this object:  $\mathrm{T_{eff}} = 936 \pm 78$ K, $log(g) =4.98 \pm 0.48$ dex, age between $0.5-10$~Gyr, M = $42.84 \pm 26.06 ~\mathrm{M_{jup}}$ and R = $0.94 \pm 0.15 ~\mathrm{R_{jup}}$.  \citet{koen2004search} and \citet{radigan2014strong} did not find any significant variability for this object with ground-based near-infrared monitoring. On the other hand, \citet{buenzli2014brown} found a tentative variability for 2M1624+0029 with HST/WFC3 data in the water band. J1624+0029 was found in the variable areas in 5 out of 12 our index-index plots, indicating that it is a non-variable candidate.

\subsubsection{2MASS J15462718--3325111}\label{2MASS J15462718-3325111}

J1546--3325 is a T5.5 dwarf \citep{burgasser2002spectra}.  \citet{smart2013nparsec} derived some physical parameters for this object: $\mathrm{T_{eff}} = 1002 \pm 84$ K, $log(g) =5.00 \pm 0.48$ dex, age between $0.5-10$ Gyr, M = $44.13 \pm 26.34 ~\mathrm{M_{jup}}$ y R = $0.94 \pm 0.16 ~\mathrm{R_{jup}}$. \citet{koen2004search} found no variability in the near-infrared after 4.4 hr of photometric monitoring. Similarly, \citet{radigan2014strong} found no significant variability for J1516-3325 after 3.8~hr of $J$-band monitoring. J1546--3325 was found in the variable areas in 3 out of 12 our index-index plots, indicating that it is a non-variable candidate.

\subsubsection{2MASSI J0243   137--245329}\label{2MASSI J0243137-245329}

J0243--2453 is a T6 dwarf \citep{burgasser2002spectra}. \citet{filippazzo2015fundamental} provided physical parameters for this object: $\mathrm{T_{eff}} = 972 \pm 83$ K, $log(g) =4.99 \pm 0.47$ dex, age around $0.5-10$ Gyr, M= $43.59 \pm 26.41~\mathrm{M_{jup}}$ and R = $0.94 \pm 0.16~\mathrm{R_{jup}}$. \citet{buenzli2014brown} found no variability $\sim 40$ minutes near-infrared monitoring. Similarly, \citet{radigan2014strong} did not find any significant variability in the $J$-band after 3.7 hr of monitoring. J0243--24532 was found in the variable areas in 4 out of 12 our index-index plots, indicating that it is a non-variable candidate.

\subsubsection{2MASS J18283572-4849046}\label{2MASS J18283572-4849046}

J1828--4849 is a T5.5 brown dwarf \citep{burgasser2000discovery}. \cite{faherty2012brown}  provided the most relevant physical parameters for this object: $\mathrm{T_{eff}} = 1060 \pm 103$~K, $log(g) =5.01 \pm 0.48$~dex, age between $0.5-10$~Gyr, M= $ 45.21 \pm 27.07~\mathrm{M_{jup}}$ and R $= 0.95 \pm 0.16~\mathrm{R_{jup}}$. Our spectral indices did not select J1828--4849 as a variable candidate, since it appeared only in 7 out of 12 variable areas. Nevertheless, \citet{radigan2014strong} found significant variability for the object, with a variability amplitude of $0.9 \pm 0.1 \%$, with a 5 hr rotational period. However, they found that the J1828--4849 light curve followed the same trend of a calibration star. Thus, it is possible that the variability measured for J1828--4849 is not intrinsic. \cite{clarke2008search} found an attenuation of 0.04 mag for the J1828--4849 light curve. Nevertheless, they found a strong correlation with the airmass, suggesting that the attenuation measured could be contamination from the control star they used.  J1828--4849 was found in the variable areas in 6 out of 12 our index-index plots, indicating that it is a non-variable candidate.

\subsubsection{SDSSp J111010+011613}\label{SDSSp J111010+011613}

J1110+0116 \citep{geballe2002toward} is a T5.5 brown dwarf, member of the 150 Myr old AB-Doradus moving group \citep{Gagne2015}.  \citet{vos2018variability} monitored it for 9 hr of observation  with Spitzer in the [4.5] channel, detecting no variability. For periods $<$18~hr, they place an upper limit of 1.25\% on the peak-to-peak variability amplitude of this object.  J1110+0116 was found in the variable areas in 7 out of 12 our index-index plots, indicating that it is a non-variable candidate.

\subsection{New Variable Candidates}\label{sec:new_variable_candidates}

The following targets are selected as new variable candidates by our spectral indices.

\subsubsection{2MASSI J0727182+171001}

J0727+1710 was classified as a T7.0 dwarf by \citep{burgasser2002spectra}. This object was photometrically monitored in a search for variability with the 1.6~m telescope of  the Observatoire du Mont-Megantic by \citet{2013ApJ...767...61G}, finding variability Peak-to-Peak Amplitudes $< 13$ mag. \citet{2016ApJ...826...73P} conducted an optical survey with the Keck telescopes looking for $H\alpha$ emission at  6563 $\text{\AA}$, which is possibly indicative of magnetic or auroral activity in brown dwarf atmospheres sometimes resulting in rotational variability. However, J0727+1710 did not  demonstrate $H\alpha$ emission, with values of $log(L_{H\alpha}/L_{bol}) < -5.9$. J0727+1710 was found in the variable areas in 11 out of 12 our index-index plots, indicating is a variable candidate.

\subsubsection{2MASSI J1217110--031113}

J1217--0311 is a T7.5 brown dwarf  \citep{burgasser2000discovery}. \citet{2014ApJ...793...75R} did not detect significant photometric variability during 3.18~hr of monitoring. This object does not show   $H\alpha$ emission \citep{2016ApJ...826...73P}, with values of $log(L_{H\alpha}/L_{bol}) < -5.9$. J1217--031 was found in the variable areas in 11 out of 12 our index-index plots, indicating that it is a variable candidate.

\subsubsection{2MASSI J1553022+153236}

J1553+1532  was discovered by \citet{burgasser2002spectra}, and classified as T6.0 by \cite{burgasser2010spex}.  \citet{dupuy2012hawaii} discovered that it is actually is a 0.349'' binary pair of T7 and T6.5 brown dwarfs resolved with HST with an estimated mass ratio of $0.9 \pm 0.02$ and a bolometric luminosity ratio of $0.31 \pm 0.12$ \citep{burgasser2006method}.  J1553+1532 was found in the variable areas in 11 out of 12 our index-index plots, indicating that it is a variable candidate.

\subsubsection{2MASS J00345157+0523050}
J0034+0523 is a T6.5 brown dwarf discovered by \citet{burgasser20042mass}. Its spectral analysis revealed that this object has a depressed K-band peak, indicating potentially a metal-poor atmosphere. Its spectrum is similar to the unusual T6 brown dwarf, J0937+2931 \citep{2002ApJ...564..421B}, since both objects have spectral indicators of a high pressure photosphere, which can either be indicative of high surface gravity or a metal poor atmosphere. J0034+0523 has a very blue near-infrared color $J-K = -0.93 \pm 0.03$, and it is one of the fastest rotating brown dwarfs found to date with a maximum rotation period of 1.4 hours and $v sin i = 90 \pm 2 km s^{-1}$ \citep{2021ApJS..257...45H}. J0034+0523 was found in the variable areas in 11 out of 12 our index-index plots, indicating that it is a variable candidate.

\subsubsection{2MASS J16150413+1340079}

J1615+1340 was not monitored for variability studies in previous works. J1615+1340 was discovered by \citet{Looper2007}, who classified it as a T7.0 spectral type brown dwarf. J1615+1340 was found in the variable areas in 12 out of 12 our index-index plots, indicating that it is a variable candidate.

\subsubsection{SDSS J175805.46+463311.9}
J1758+4633 was not monitored for variability studies in previous works. It was discovered by \citet{2002ApJ...564..466G}, and classified it as a T7.0 brown dwarf. J1758+4633 shows also signs of low surface gravity  \citep{2006ApJ...639.1095B}. J1758+4633 was found in the variable areas in 11 out of 12 our index-index plots, indicating that it might is variable candidate.

\subsubsection{SDSSp J134646.45--003150.4}
J1346--0031 was not monitored for variability studies in previous works. J1346--0031 was discovered by \citet{2000ApJ...531L..57B} and classified it as a T6.0 spectral type brown dwarf. This object does not have any noticeable magnetic activity, with H$\alpha$ emission values of $log(L_{H\alpha}/L_{bol}) < -5.2$. J1346--0031 was found in the variable areas in 11 out of 12 our index-index plots, indicating that it is a variable candidate.

\section{Robustness of the Index Method}\label{sec:robustness}

\begin{figure}
    \centering
    \includegraphics[width=1\linewidth]{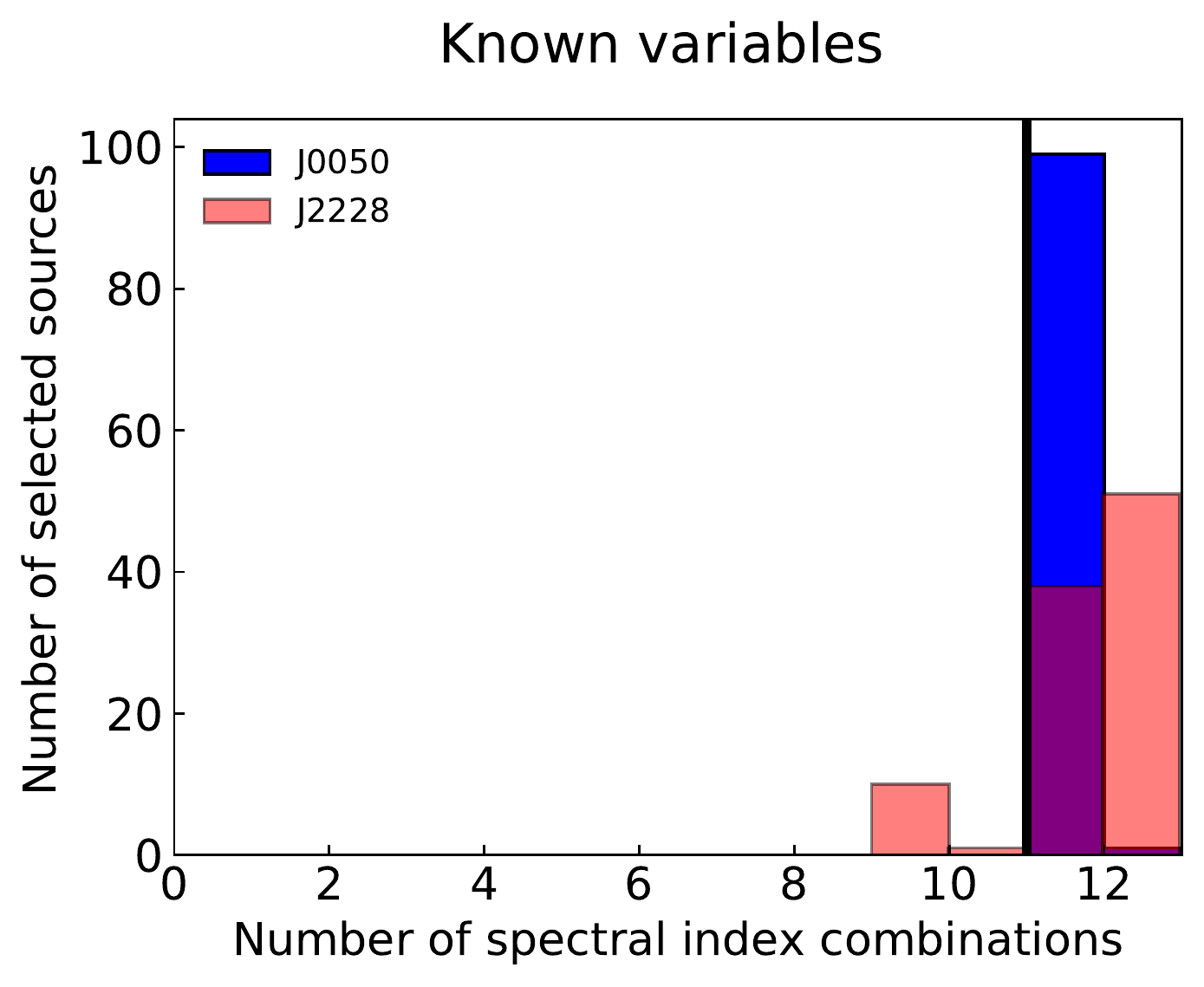}
    \caption{Histogram of the number of variable zones in which the variable synthetic spectra generated with the SpeX spectra are found. Most of the synthetic spectra generated from the SpeX spectra of 2M2228 and J0050--3322 are in more than 11 variable zones in the index plots.}
    \label{fig:histogram_variables}
\end{figure}

Since we have a limited sample of mid and late-T with variability information, we aim at testing the method for a larger sample of spectra. 
To verify the robustness of our method, we generated synthetic spectra using a Monte Carlo simulation for the known variables and non-variables dwarfs that were selected as such by our spectral indices (see Sections \ref{sec:variables} and \ref{sec:non-variables}). Using these 8 objects (2 variables and 6 non-variables), we performed a similar Monte Carlo simulation as the one described in Section \ref{sec:index-method} using their IRTF/SpeX near-infrared spectra. 100 synthetic spectra were generated for each object using the Monte Carlo simulation. We calculated the six spectral indices mentioned in Table \ref{Limits_ranges_index}. Then, these spectral indices are compared with each other in the index-index plots as in Fig. \ref{index-index_comparison}. Finally, we calculated how many of the synthetically generated spectra are found in more than 11 variable regions within the index-index plots for the variable and the non-variable spectra. In Figure \ref{fig:histogram_variables}, we present the histogram showing in how many variability regions the synthetic spectra created using the SpeX spectra of the two variable objects (2M2228 and 2M0050--3322) are found. For J0050--3322, 100\%  of the synthetic spectra fall the variable regions in more than 11 index-index plots. For J2228, 89\% of the synthetic spectra are found in more than 11 variable regions in the index-index plots.

\begin{figure}
\centering
    \includegraphics[width=1\linewidth]{ 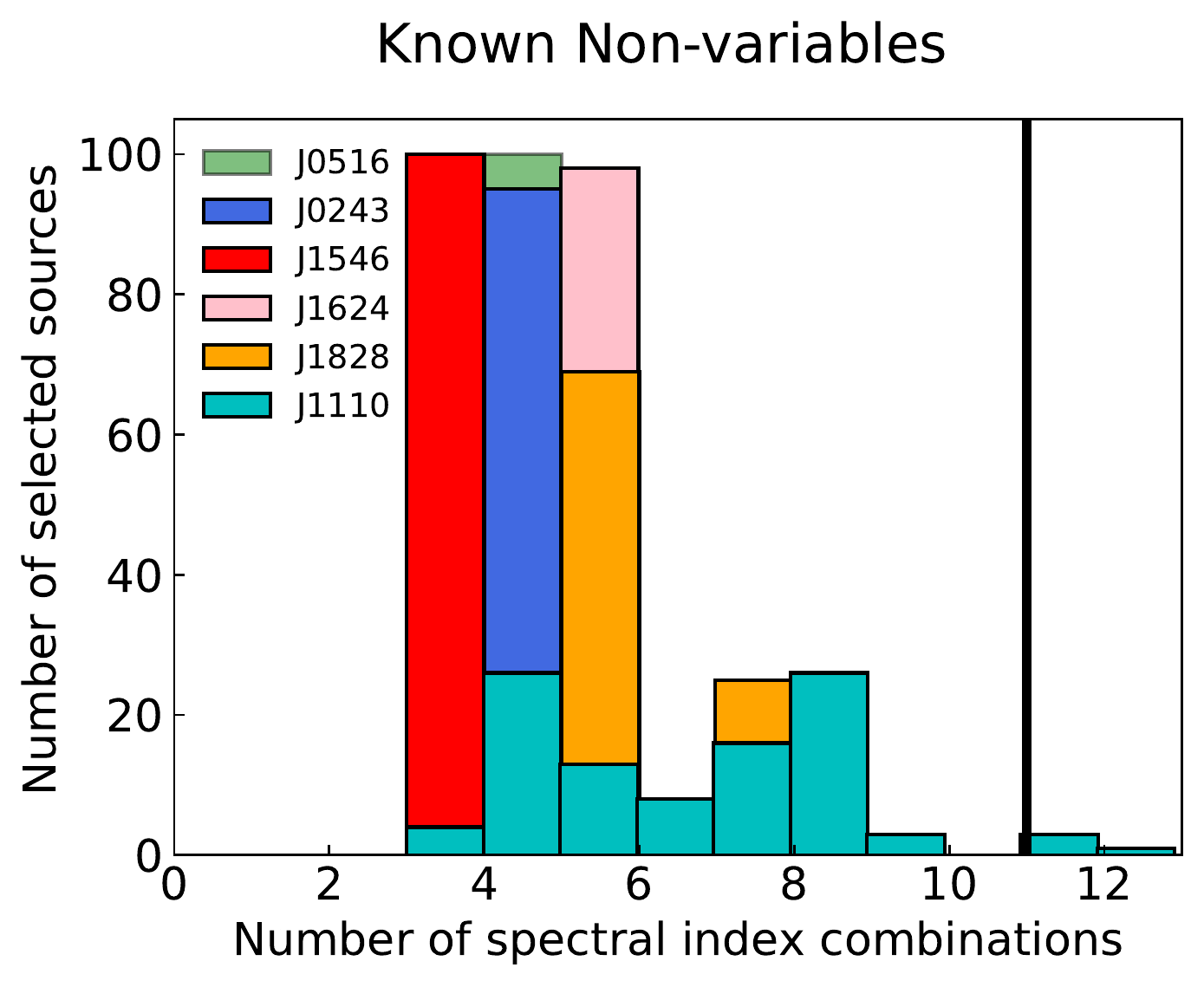}
    \caption{Histogram of the number of times in which non-variable synthetic spectra generated using the SpeX spectra fall. Only J1110+0116 have a few spectra in more than 11 out of 12 regions.}
    \label{fig:histogram_novariables}
\end{figure}

On the other hand, in Figure \ref{fig:histogram_novariables}, we show in how many variable regions the synthetic spectra generated using the SpeX spectra of the known non-variable objects are found. The majority of  the objects fulfill the non-variable condition since 100\% of their spectra are found in less than 11 variable regions.

As mentioned in Section \ref{ULAS J141623.94 + 134836.3}, J1416+1348B is found in 10 index-index plots, falling in the non-variable category. J1416+1348B is a blue object (\citealt{burgasser2010spex}, see Fig. \ref{fig:color-mag}), thus its spectrum and colors differ from the standard mid- to late-T spectral sequence. Thus, J1416+1348B might be a contaminant in our poll of variable/non-variable candidates. J1416+1348B has an upper limit on variability \citep{metchev2015weather}. 

\begin{figure}
\centering
    \includegraphics[width=1\linewidth]{ 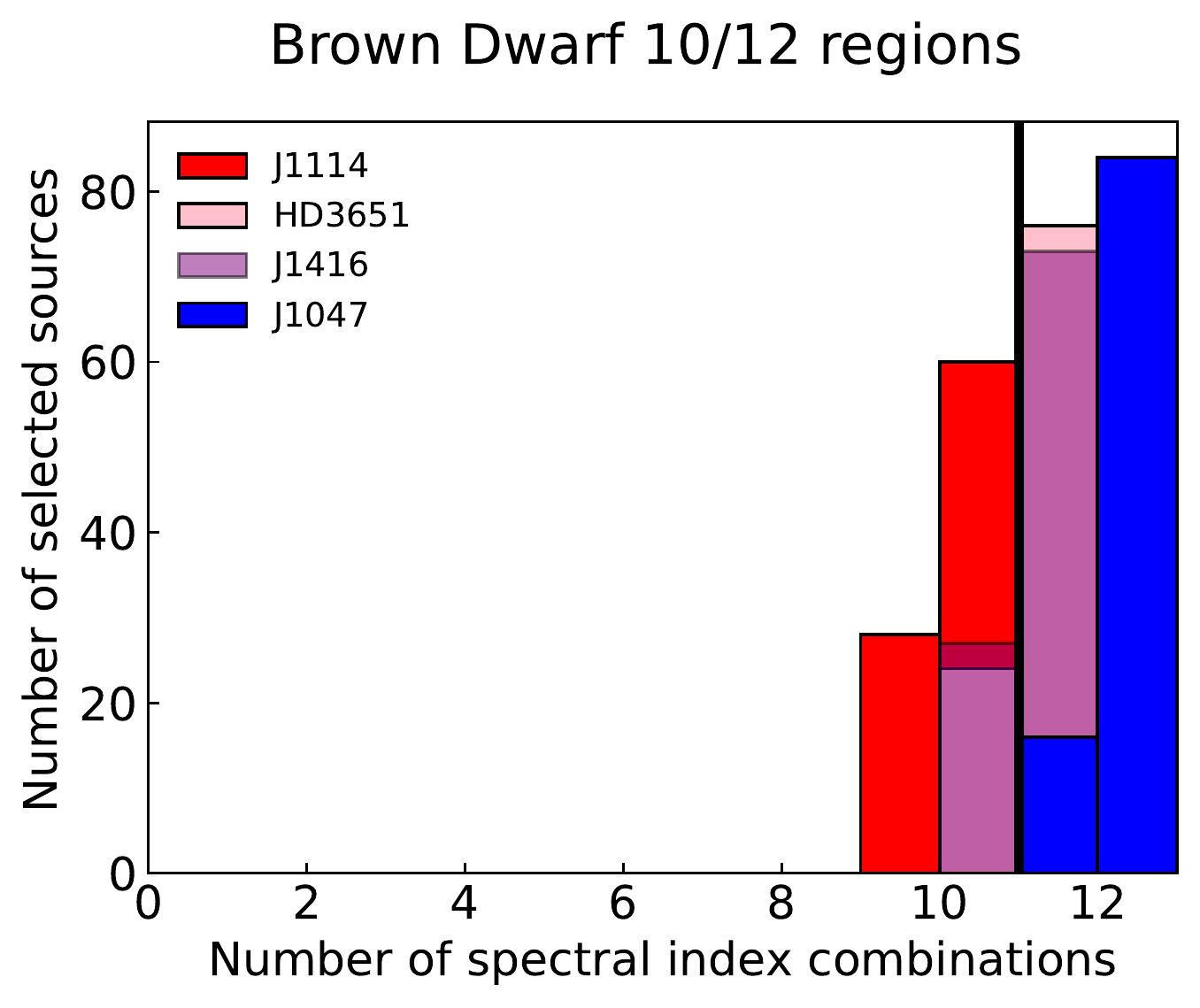}
    \caption{Histogram of the number of times the synthetic spectra of brown dwarfs that are in the 10/12 limit of our variability regions.}
    \label{fig:histogram_medium_test}
\end{figure}

We consider if objects that found in 10/12 areas should be considered also variable candidates. To test how robust this limit is, using a Monte Carlo simulation as in Section \ref{sec:robustness}, we generated 100 synthetic spectra for each of the four objects that found in 10/12 areas (J1114, HD3651, J1416B and J1047), and we estimated how many of the newly generated 100 synthetic spectra found in more than 11 out of 12 areas. As shown in Fig. \ref{fig:histogram_medium_test}, with the exception of J1114--2618 (found only 12/100 as candidate variable), most of the synthetically generated spectra from  the other three objects found in more than 11 of 12 areas. HD3651 76/100 variable spectra, J1416+1348B 73/100 variable spectra, J1047+2124 100/100 variable spectra.

In conclusion, to acknowledge this finding, the objects that found in 10 out of 12 variability areas might be considered ``weak variable candidates".

\section{Discussion}\label{Sec:discussion}

\subsection{Recovery Rates}\label{recovery_rates}

In Table \ref{table:sample_testing}, we find that 10 objects in the sample of 26 T5.5-T7.5 dwarfs were monitored for variability studies in the literature. Three of them were found to be variable (2M2228, J0050-3322 and J1047+2124), while the other seven are not variable. The index method presented in Section \ref{sec:index-method} was able to recover two of the three known variable brown dwarfs (recovery rate of 67\%). For the known non-variable brown dwarfs, our index method was able to recover all seven known non-variables (recovery rate of 100\%), we find single false negative, J1047+2124, which would suggest a false-negative rate of 12.5\% (1/8). As mentioned in Section \ref{2MASSI J1047538+212423}, J1047+2124 is a borderline variable candidate, selected as variable by 10/12 regions, and we decided to consider it as a "weak variable candidate". Further variability monitoring campaigns to observe the new variable candidates in Section \ref{sec:new_variable_candidates} are needed to confirm the recovery rates of our index method.

\subsection{Variability Fraction}\label{comparison_surveys}

Using the index method presented in this paper (Section~\ref{sec:index-method}) over a sample of 26 mid and late-T dwarfs SpeX near-infrared spectra with spectral types between T5.5 and T7.5,  and based on the assumption that the variable candidates are indeed variable, we aim at estimating the variability fraction given by our indices, and compare it with other estimations in the literature. 

We found that 10 of the 26 objects are variable candidates. This suggests a variability fraction of $\mathrm{38^{+4}_{-30}}$\% for this spectral range. 
From the 10 variable candidates, only 2 are known variables (2M2228 and J0050--3322). We followed the same method as in \cite{Burgasser2003} to estimate the uncertainties of the variability fraction using a Poisson distribution. Considering that the total number of objects in our sample is 26, the mean of the distribution is 10 (variable candidates found by our indices), and the  minimum variable fraction of our sample is 2/26 $\sim$ 8\%. We integrated the  probability of the distribution until reaching 0.68, equivalent to 1-$\sigma$ Gaussian limits. We  estimated the variability fraction of our sample in $38^{+4}_{-30}$\% according to our indices. However, taking into account the false negative rate of the indices, the true fraction of variability could be slightly higher.

We compared this variability fraction with the results of ground-based \citep{radigan2014strong} and space-based (\citealt{buenzli2014brown}, \citealt{metchev2015weather}) surveys to understand if the variability fraction reported in this work is consistent with those. \cite{radigan2014strong} monitored a sample of 57 L, L/T and T-dwarfs in the $J$-band using the Du Pont 2.5 m telescope at Las Campanas Observatory and the Canada–France–Hawaii Telescope on Mauna
Kea. \cite{radigan2014strong} had 15 dwarfs with a spectral type later than T5.5, and reported significant variability for three of them, implying a variability fraction of $\sim$20\%. Because of the limited stability and sensitivity of the ground-based data, this 20\% should probably be considered as a lower limit  rather than an intrinsic variability fraction. Similarly, \cite{buenzli2014brown} monitored a sample of 22 L5 to T6 dwarfs with the \textit{Hubble Space Telescope/Wide Field Camera 3} in the $J$- and $H$-bands, finding six brown dwarfs with confident variability and five brown dwarfs with tentative variability. This leads to a variability fraction between 27--50\%. In addition, \cite{metchev2015weather} monitored a sample of 44 L3 to T8 dwarfs using the \textit{Spitzer} Space Telescope in the [3.6] and [4.5] channels. \cite{metchev2015weather} monitored five T-dwarfs with spectral types later than T5.5, finding significant variability  for two of them ($\sim$40\%). \cite{metchev2015weather} concluded that most of the T-dwarfs in their sample show low level variability ($>$0.2\%). The difference in variability fractions is most likely due to three factors: the difference in sensitivity between ground and space-based telescopes, the wavelength range at which each sample was observed, and the different time-scales: the \textit{Spitzer} observations were usually much longer than what is possible from the ground. The $J$-band traces deeper pressure levels of the mid to late-T atmospheres ($\sim$5~bar, \citealt{yang2016extrasolar}, see their Fig. 19) than the [3.6] and [4.5] \textit{Spitzer} bands ($\sim$0.2~bar and $\sim$0.6~bar, respectively, \citealt{yang2016extrasolar}). Given that most of the cloud deck is expected to be at pressures above  $\sim$1~bar \citep{yang2016extrasolar}, in principle, we do not expect a high variability in the \textit{Spitzer} channels. A visual representation of the atmospheric structure of a similar spectral type object, 2M0050-3322, can be found in \cite{Manjavacas2022}, their Fig. 19. Nevertheless, the sensitivity of the \textit{Spitzer} Space Telescope is higher than ground-based observatories. The variability fraction we report on this work, 38$^{+4}_{-30}$\%, is consistent to the \textit{Spitzer} variability fraction reported by \cite{metchev2015weather}.

\subsection{Physics behind the index method}\label{physics}

\begin{figure}
\vspace{0.3cm}
    \centering
    \includegraphics[width=0.99\linewidth]{ 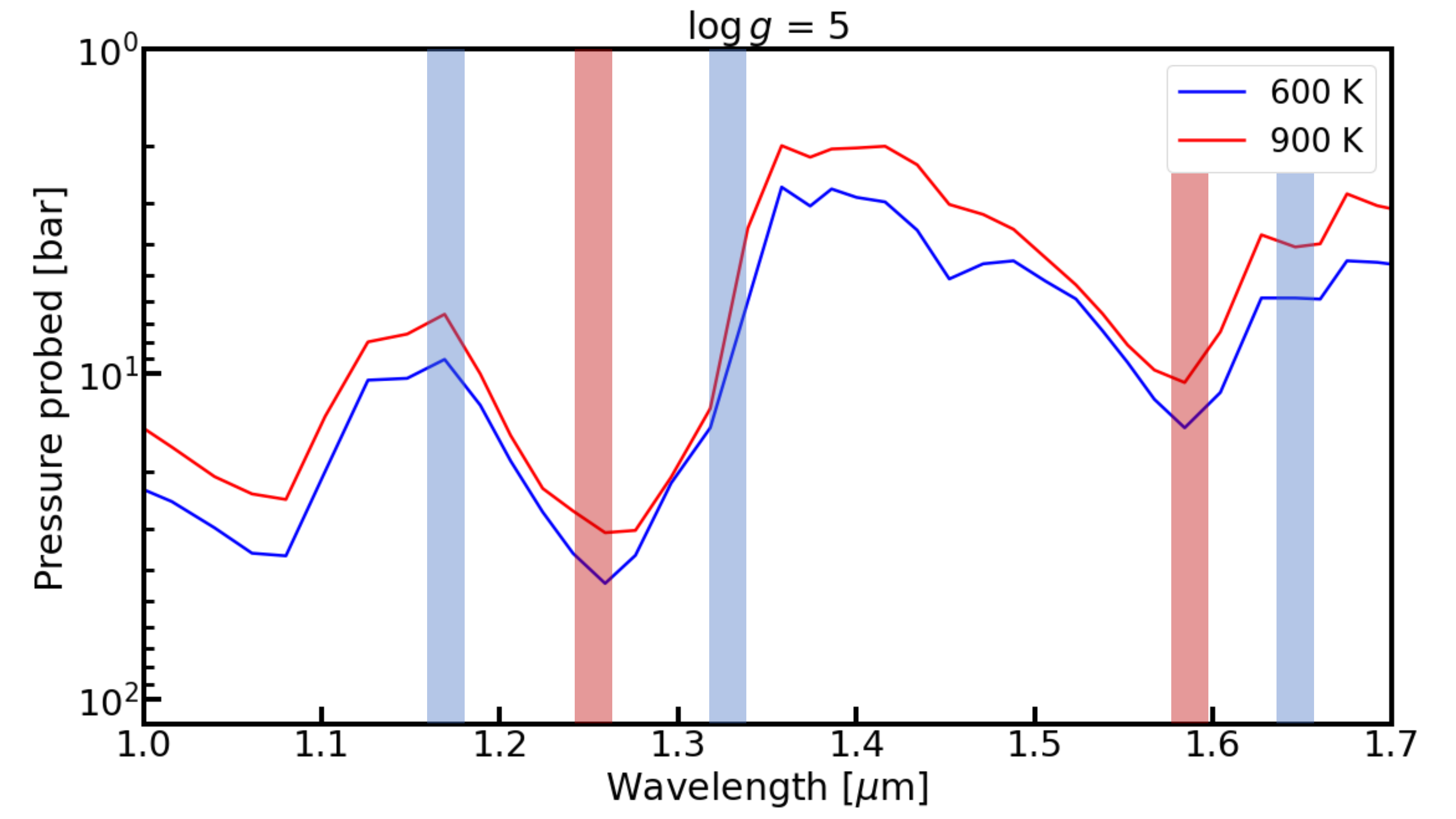}
    \caption{Pressure levels probed by the HST/WFC3 + G141 wavelength range for T-dwarfs with temperatures of 900 K (red line) and 600 K (blue line), covering the T5-T8 spectral types according to radiative-transfer models. We shadowed in red the most variable wavelength ranges of the spectra, which were used as numerator for J and H indexes (see Fig. \ref{fig:ranges_index}). We shadowed in blue the least variable wavelength ranges of the spectra, which were used as denominator for indixes J, H and J-slope}. \label{fig:pressure_levels}
\end{figure}

The spectral indices presented in this work compare the wavelength ranges of the $J$- and $H$-bands of the HST/WFC3 spectra of 2M2228 that vary the most  with the wavelength ranges that vary the least as the object rotates. The most variable ranges of the $J$- and $H$-bands are usually located around the maximum of the bands (see Fig. \ref{fig:ranges_index}), and the least variable are usually located at the wings of the bands (see Fig. \ref{fig:ranges_index}). In addition, two of our indices compare the $J$- and $H$-band flux between them. To understand the reason why certain wavelength ranges show higher flux variations than others, we use the vertical structure predicted for an object of similar spectral type and surface gravity to 2M2228 by radiative-trasfer models \citep{Saumon_Marley2008} in \cite{yang2016extrasolar}. In Fig. \ref{fig:pressure_levels} we show the pressure levels probed by HST/WFC3 and the G141 grism, similar to the contribution functions presented in \cite{yang2016extrasolar}, their Fig. 18. As seen in Fig. \ref{fig:pressure_levels}, the peaks of the $J$- and $H$-bands around 1.25~$\mu$m and 1.58~$\mu$m (shaded in red), respectively, probe the deepest layers of the atmosphere of a mid to late-T dwarf that we are able to trace. The peak of $J$-band probes the 30--40~bar level, and the $H$-band peak probes the 10--20~bar level. In the meantime,  the wings trace  upper layers: the $\sim$10~bar level is traced with wavelengths around 1.17~$\mu$m and 1.32~$\mu$m for the $J$-band, and the $\sim$6~bar level is traced with wavelengths around  1.65~$\mu$m in the $H$-band (see Fig. \ref{fig:pressure_levels}, shaded in blue). As discussed in Section \ref{sec:index-method}, these the wavelength ranges covered by the wings  barely change during the rotational period of 2M2228. 

In addition, the condensate mixing ratio (mole fraction) of the clouds present in 2M2228's atmosphere ($\mathrm{Al_{2}O_{3}}$, Fe, and $\mathrm{Mg_{2}SiO_{4}}$ clouds) is predicted to decrease with decreasing pressure (see \citealt{yang2016extrasolar}, their Fig. 19). Thus, as we probe higher levels in the atmosphere of 2M2228, we expect to have less dense clouds, and therefore potentially less variability too. This scenario would explain why the variability observed in the wings of the $J$- and $H$-band is smaller than in their respective peaks. Finally, the high signal-to-noise of the HST/WFC3 near-infrared spectra is enough to quantify the differences in variability amplitude, and thus to design the spectral indices that allowed us to identify new $>$T5.5 variable candidates (see Section \ref{Sec:Candidate-variables}) with a single-epoch near-infrared spectrum.

\subsection{Top-of-the-atmosphere features}\label{atmospheric_characteristics}

In contrast to early-T dwarfs, mid- and late-T dwarfs show in general relatively low variability amplitudes \citep{radigan2014strong, metchev2015weather}. Variability surveys from the ground and space-based facilities have not found variability amplitudes above the 2\% for mid- and late-T dwarfs, with the only exception  of Ross~458~c  \citep{manjavacas2019ROSS}. Nevertheless, \cite{metchev2015weather} showed that most mid- and late-T dwarfs do show low-level variability ($>$0.4\%). Even though few mid- and late-T dwarfs have been monitored for spectrophotometric variability with space-based facilities that provide high signal-to-noise data, they are expected to show bands and spots as Solar System planets do, according to General Circulation models (GCMs) \citep{2021MNRAS.502..678T}. GCMs provide instantaneous model maps for the top-of-the atmosphere thermal flux, and predict the most likely light curve for such an object. The variability amplitude predicted for a T-dwarf of a similar spectral type to 2M2228 (2M0050--3322, \citealt{Manjavacas2022}) is about $\sim$1\% for an edge-on object. The predicted variability amplitude decreases to about 0.2\% for a pole-on object. GCMs predict a band near the equator driven by the radiative feedback of MnS and $\mathrm{Na_{2}S}$ clouds. A more extended sample of mid- and late-T dwarfs with high signal-to-noise spectrophotometric monitoring data from space-based facilities would help to constrain the properties and dynamics of mid- and late-T dwarfs, and would allow to confirm the prediction of GCMs.

\subsection{Analogs to Cool Directly-Imaged Exoplanets}\label{exo_analogs}

Brown dwarfs are analogs to directly-imaged exoplanets, as they share effective temperatures, colors, and sometimes surface gravities and masses (e.g. \citealt{2016ApJS..225...10F}). This is the case of the 51~Eridani~b directly-imaged exoplanet (600--750~K, \citealt{2015Sci...350...64M}), HD 19467 b (1040$\pm$40~K, \citealt{2019ApJ...873...83W}),  Gl 758 b (600--750~K, \citealt{2018AJ....155..159B}), or HD 4113C (500--600~K, \citealt{2018A&A...614A..16C}), which show significant variability as in the case of HR8799bc   \citet{biller2021high}, among others. Thus, a comprehensive atmospheric characterization of the technically easier to observe isolated T-dwarfs might shed light on the atmospheric structures of cool directly-imaged exoplanets \citep{apai2017exploring}. High signal-to-noise spectro-photometric time-resolved data provide information about the variability introduced at different pressure levels of the atmospheres of these objects, which allow us to understand at which pressure levels most of the heterogeneous clouds introducing the measured variability are found, using radiative-transfer models (e.g. \citealt{yang2016extrasolar}). Continued monitoring of few rotational periods of these objects allow us to produce a map of its surface and reconstruct which atmospheric structures (band and/or spots) have most likely produced its light curve at different pressure levels. 

Mapping Codes  like \textit{Stratos} \citep{2013ApJ...768..121A}, \textit{Aeolus} \citep{2015ApJ...814...65K} or \textit{Starry} \citep{2019AJ....157...64L} allow us to generate 2D maps of a given light curve expanding over 2-3 rotational periods. Such a detailed atmospheric characterization is yet technically challenging for directly-imaged exoplanets \citep{apai2016high}. Nevertheless, with the advent of the first scientific data from the just launched \textit{James Webb Space Telescope}, we will be able to characterize more cool directly-imaged exoplanets. The spectral indices presented in this paper will be of a special importance, since they will allow the community to pre-select the most likely variable T-dwarf type-like directly imaged exoplanets to be monitored for spectro-photometric variability. Since until now no systematic method to pre-identify variable T-dwarfs exists, these spectral indices will very likely save hours of precious telescope time.

\section{Conclusions}\label{Sec:conclusions}

\begin{enumerate}

    \item We designed  a set of novel near-infrared spectral indices to find variable candidate mid- and late-T ($>$T5.5) dwarfs using a single-epoch near infrared spectrum. To design these novel spectral indices, we used \textit{HST/WFC3} near-infrared, high signal-to-noise, time-resolved spectra of the  T6.5 variable 2M2228-4310 published in \cite{buenzli2012vertical}. 
    
    \item We designed six new spectral indices to find potential variable mid- and late-T dwarfs. With those six indices, we created twelve index-index plots which allow us to segregate the variable and non-variable spectra of 2M2228, and  to define areas within those plots where we expect to find variable mid- and late-T dwarfs.
    
    \item Using a Monte Carlo simulation of the most variable and least variable spectra of 2M2228, we refined the  areas in which variable mid- and late-T dwarfs should be find within the index-index plots.
    
    \item Using the synthetically generated spectra of 2M2228, we flagged  the objects that were found in more than 11 out of 12 variable areas as variable candidates. Objects that were found in 10 out of 12 variable areas might be flagged as weak candidate variables, since some of them have been found to be marginally variable.
    
    \item Using the newly-generated near-infrared spectral indices, we identified potential mid- and late-T variable brown dwarfs using their single-epoch spectra. We used all the mid- and late-T dwarf near-infrared spectra available at the SpeX/IRTF spectral library with spectral types between T5.5 and T7.5.  We reported a variability fraction of $38^{+4}_{-30}$\%  for our late-T dwarf sample assuming that the variable candidates selected are indeed variable. The variability fraction reported agrees with the variability fraction provided by \cite{metchev2015weather} for the same range of spectral types.

    \item From the 10 variable candidates selected by the spectral indices, three were known variables (2M0050--3322, 2M2228-4310 and J1047+2124). Of the objects selected as non-variable, seven are known non-variables (J1416+1348B, J0501--0445, J1624+0029, 2MASS~J1546-3325,  J1024--2453, J1828-4849, and J1110+0116). Using the index method we are able to recover two of the three known variables (recovery rate of 67\%), and all the known non-variables (recovery rate 100\%). J1047+2124 is a false negative, which would suggest a false negative rate of ~12.5\%.

    \item We reported eight new mid- and late-T variable candidates, which will need to be confirmed with near-infrared photometric monitoring follow-up.
    
    \item To test the robustness of our indices, we produced 100 synthetic spectra using the SpeX/IRTF spectra of the known variable and non-variable mid- and late-T dwarfs with a Monte Carlo simulation. We calculated their spectral indices and place them in the index-index plots. We concluded that the great majority of the synthetic spectra created for the known variables were also classified as variables by our indices. Similarly,  the synthetic spectra created for the known non-variables were classified as non-variable by our indices, with the exception of  J1416+1348B.
    
    \item The indices presented in this work measure the difference in variability at the peak of the $J$- and $H$-bands, and at their sides of the bands. The peak of the $J$- and $H$-bands trace the deeper layers of the atmosphere of mid- and late-T dwarfs (30-40~bar, and 10--20~bar, respectively), and the sides of the bands trace upper levels ($\sim$10~bar and $\sim$6~bar, respectively). Thus, if a difference in heterogeneous cloud coverage exist at these different pressure levels, the variability should change among those, affecting slightly the flux at the peaks of the $J$- and $H$-bands, and at their sides.
    
    \item Mid- and late-T dwarfs are analogs to cool directly-imaged exoplanets in terms of temperatures and colors. Thus, the spectral indices presented in this work could help to pre-select the most-likely variable exoplanet candidates in the near future, to be monitored in spectro-photometric variability studies with the \textit{James Webb Space Telescope} or the 40~m class telescopes scheduled to start observations during this decade.
    
\end{enumerate}
\vspace{2mm}
\section{Acknowledgments}
We thank the RECA\footnote{\url{https://www.astroreca.org/en/}} (Red Estudiantes Colombianos en Astronom\'ia) Summer Internship Program for providing an avenue to initiate the discussions that led to this work.


\bibliography{sample631}{}
\bibliographystyle{aasjournal}



\end{document}